\documentclass[aps,prd,preprintnumbers,showpacs,superscriptaddress,nofootinbib,amsmath,amssymb,floats,floatfix,showkeys,notitlepage,longbibliography]{revtex4-1}
\usepackage{comment}
\usepackage{graphicx}
\usepackage{subfigure}
\usepackage{palatino}
\usepackage[commandnameprefix=always]{changes}
\usepackage{hyperref}
\hypersetup{colorlinks=true,linkcolor=red,urlcolor=red,citecolor=red}
\usepackage[toc,page]{appendix}
\usepackage[normalem]{ulem}

\usepackage{lipsum}
\usepackage{graphicx}
\usepackage{subfigure}
\usepackage{palatino}
\usepackage{float}
\usepackage{sans}
\usepackage{adjustbox}
\usepackage{latexsym}
\usepackage{amsmath}
\usepackage{amssymb}
\usepackage{amsfonts}
\usepackage{dcolumn}
\usepackage{bm}
\usepackage{tikz}
\usepackage{bigints}
\usepackage{array,tabularx,multirow,booktabs}
\usepackage[tracking=true]{microtype}
\SetTracking{}{500}
\SetTracking{encoding={*}, shape=sc}{40}
\allowdisplaybreaks
\usepackage{adjustbox}
\usepackage{latexsym}
\usepackage{amsmath}
\usepackage{amssymb}
\usepackage{amsfonts}
\usepackage{dcolumn}
\usepackage{bm}
\usepackage{tikz}
\usepackage{bigints}
\usepackage{array,tabularx,multirow,booktabs}
\usepackage[tracking=true]{microtype}
\usepackage{color}
\allowdisplaybreaks

\begin{document}

\title{Kramer's Escape Rate and Phase Transition Dynamics in AdS Black Holes}

\author{Mohammad Ali S. Afshar}
\email{m.a.s.afshar@gmail.com}
\affiliation{Department of Physics, Faculty of Basic Sciences,
University of Mazandaran\\
P. O. Box 47416-95447, Babolsar, Iran}
\affiliation{Canadian Quantum Research Center, 204-3002 32 Ave, Vernon, BC V1T 2L7, Canada}

\author{Saeed Noori Gashti}
\email{saeed.noorigashti@stu.umz.ac.ir}
\affiliation{School of Physics, Damghan University, P. O. Box 3671641167, Damghan, Iran}

\author{Mohammad Reza Alipour}
\email{mr.alipour@stu.umz.ac.ir}
\affiliation{School of Physics, Damghan University, P. O. Box 3671641167, Damghan, Iran}
\affiliation{Department of Physics, Faculty of Basic Sciences,
University of Mazandaran\\
P. O. Box 47416-95447, Babolsar, Iran}

\author{Jafar Sadeghi}
\email{pouriya@ipm.ir}
\affiliation{Department of Physics, Faculty of Basic Sciences,
University of Mazandaran\\
P. O. Box 47416-95447, Babolsar, Iran}
\affiliation{Canadian Quantum Research Center, 204-3002 32 Ave, Vernon, BC V1T 2L7, Canada}

\vspace{1.5cm}\begin{abstract}
Traditional static methods in phase transition studies, such as the swallowtail bifurcation diagram, provide good insights into the thermodynamics of black holes. However, they practically lose sight of the dynamic aspects and temporal sequence of events. The Kramers escape rate, central to our research, offers a somewhat dynamic approach to phase transition. We examine the free energy landscapes for black holes under the influence of 'dark' and 'stringy+dark' structures, assessing how additional parameters affect the escape rates and dynamics of the transition during the first-order phase transition from small to large black holes. In our analysis, we consider the escape rate as a function of the black hole radius and study its variations. We will observe that, on one hand, the escape rate well represents our assumption based on the movement from zero, increasing to a maximum point, and then decreasing back to zero as reactive structures become active during the phase transition interval. However, the critical point in this method is the encounter with a specific and distinct point. This is where the diagram of the direct process (escape rate from small to large black holes) intersects with the reverse process (large to small black holes), becoming equally probable (contact point). The point, which seems improbable at the onset of the phase transition or very negligible, gains more significance as the process progresses. This increase indicates the dominance of a region where the escape rate from larger black holes to smaller ones prevails. The predominance of the reverse process, which increases as we approach the end of the process and is necessarily accompanied by a variation in radius, may be considered as a natural reaction of the black hole against the 'phase change' action. A reaction which attempting to prevent any uncontrolled radial growth that could jeopardize the stability of the black hole.
\end{abstract}

\date{\today}



\maketitle
\section{Introduction}
From ancient times to the present, the existence of similar behavioral patterns and recurring elements in various physical systems, ranging from subatomic to cosmic scales, has been abundantly observed. This could be interpreted as a mysterious sign of a fractal structure inherent in the essence and fabric of nature. However, it can also be considered an intentionally hidden shortcut, ingeniously and knowingly embedded in the labyrinth of nature to guide us step by step in deciphering complexities that we cannot grasp all at once. Among the structures that have benefited most from the analogy of behavioral patterns for theoretical modeling, due to their inaccessibility for practical study, are black holes. The comparison of the gravitational field treatment of a black hole and its surrounding world, as a system and environment, with the behavior of a thermodynamic ensemble is indeed one of the most valuable achievements of this modeling. A resemblance that has led to black hole thermodynamics becoming a pioneering and fundamental branch in the evolution of our understanding not only of the structural nature of black holes, but also of the cosmos itself. If we were to highlight the three principal keys that paved the way for the application of thermodynamics and statistical mechanics to the study of black holes, we would undoubtedly mention Hawking temperature, Bekenstein-Hawking entropy, and the conversion of the cosmological constant to pressure. These concepts led to the formulation of the four fundamental laws of black hole thermodynamics \cite{1,2,3,4}. With a slight indulgence, one might also add the exchange and conversion of enthalpy to mass to these golden keys \cite{5}.The increasing study of black hole models based on thermodynamics and statistical mechanics has gradually revealed that, apparently, the best model that aligns most closely with current data and computational capabilities, and environmental conditions for studying black hole behavior, is the Van der Waals model (VdW). For instance, in canonical ensembles, a black hole phase transition was found in the background of charged rotating AdS black holes, reminiscent of the liquid-to-gas phase transition of the (VdW) fluid \cite{6,7,8}. It was also observed that there is a significant resemblance between the (VdW) liquid-gas system and the charged black hole in the extended phase space \cite{9,10}, or another study showed the critical behavior of charged black hole in fixed charge ensemble coincides exactly with that of (VdW) liquid-gas system \cite{11}. Undoubtedly, phase transition is one of the most fundamental phenomena in classical thermodynamics, which, through extensive study of various structures of matter, has led to a profound understanding of the true behavior of systems in the past. The question that quickly arose was whether, given the (VdW)-like thermodynamic behavior of black holes, these structures, considering their quantum-relativistic structural pattern, could exhibit quasi-material behavior and phase transitions. And if so, what could be the cause of the initiation of this process? The answer to the first part of the question was subsequently obtained, as it was determined that there exists a Hawking-Page phase transition between a large stable black hole and pure thermal radiation \cite{12}. Of course, numerous other studies were conducted that even confirmed the interesting phase transition behaviors for black holes \cite{a,b,c,d,e,f,g,h, i,j,k,m}. However, the second part of the question remains unclear to us until now.\\\\
Although the temperature variation graph of a system can itself be a reliable criterion for phase transition detection, one of the tools that can be used to study the behavioral range of thermodynamic phase transition with greater precision is the free energy landscape. Free energy landscape provides a way to visualize and understand how systems, transition between different states over time. The dependence of free energy on temperature causes temperature changes to significantly alter the free energy landscape. Consequently, studying these apparent changes can provide a better understanding of the system's behavior and the formation of transitions between different phase states. In thermodynamics, free energy is rewritten in various forms depending on the system's conditions. However, since the interplay between temperature and pressure is often of primary concern in black hole structures, and since enthalpy offers a better and simpler description of energy transfer at a specific pressure, the Gibbs free energy function is a better candidate among the various definitions of free energy for studying black holes. But the point that was hidden in the study of the phase transition behavior of black holes using Gibbs free energy was that, apparently, in this form of study, no attention was paid to the rate and mode of transition from one state to another. In other words, in these studies, we did not address the following points  in the phase transition process from small to large black holes: At what rate do these processes occur? Is the rate of transition constant across different temperature ranges? Does the initial state (small black hole) completely vanish at the end of the transition? Also, could there be hidden points in the leakage from the small to the large black hole that we are unaware of? These and similar questions that address the dynamics of phase transitions provide important information about the phase transitions of black holes, which apparently cannot be obtained in the initial analyses of Gibbs free energy. It seems that we were considering the phase transition and its results in a static manner, only at the end points and equilibrium points of  Gibbs. Recent studies have observed that phase transition under thermal fluctuations can be examined through the Fokker-Planck equation in non-equilibrium physics, based on the free energy landscape, particularly the Gibbs form. Accordingly, it has been observed that in black holes, a transition occurs from small to large black holes and vice versa, influenced by temperature and the height of the potential barrier in the free energy landscape \cite{17,18,19,20}. According to the studies done, it seemed that the Gibbs free energy is in the form of an effective potential it plays the role of a driving force that leads to the phase transition of the black hole \cite{21}. Based on this, a special focus was formed on the Gibbs energy landscape, which, according to the structure of the Fokker-Planck equation, tried to investigate some kinetics of the phase transition by calculating the mean first passage time in several black holes\cite{22,23,24,25,26,27,28,29,30,31,32,33,34,35,36,37,38,39,40,41,42,43,44}. Of course, other efforts were also made in this direction based on the rewriting of free energy in terms of Landau free energy\cite{45,46} or thermal potential \cite{47,48,49,50}.\\
One might pose the question: How has the connection between stochastic-form equations and phase transitions, as well as thermodynamic equations, been established? In response, it should be noted that since some thermodynamic processes in the system are driven by stochastic fluctuations, it would be logical to employ the method of analyzing non-equilibrium stochastic processes to study the occurrence of inherently non-equilibrium thermodynamic processes. Building on this foundation, in this paper, we consider two black holes influenced by dark energy, namely: $\bullet$ Non-Linear Magnetic-Charged AdS black hole surrounded by Quintessence, in the background of Perfect Fluid Dark matter (NLM-C-Q-PFD) \cite{51}. $\bullet$ 4D AdS Einstein-Gauss-Bonnet-Yang-Mills Black hole with a Cloud of Strings (EGB-YM-CS) \cite{52}.\\
Initially, we calculate the various forms of free energy mentioned and demonstrate that all three forms will exhibit nearly identical behavior in the canonical ensemble. Subsequently, using Kramer’s escape rate method, we will examine the characteristics of their first-order phase transition rate in terms of variations in radius, pressure, or temperature. We will then compare our results with those of black holes studied using the same method  \cite{47,48,49,50} to determine the influence of the added parameters. The present paper is organized as follows: In Section 2, we explore the theoretical preliminaries to thoroughly study and introduce various concepts, such as Kramer’s escape rate and different forms of the free energy landscape. This includes the Gibbs free energy landscape, Landau free energy, and thermal potential. In Section 3, we conduct a behavioral comparison of the free energy landscape and Kramer’s escape rate for (NLM-C-Q-PFD) AdS black hole. Section 4 presents an accurate calculation for the (NLM-C-Q-PFD) AdS black hole. In Section 5, we fully analyze the concepts discussed in the previous two sections. Finally, in Section 6, we comprehensively present the results of our work.
\section{ Theoretical preliminaris }
\subsection{Kramer’s escape rate}
The study of Brownian particle leakage or escape under thermal fluctuations from a stable or metastable potential, known as the Kramers problem, is a fascinating topic that has garnered attention across most branches of physics \cite{50}. This subject, in addition to the diverse results it has yielded throughout its study, has led to the calculation of a relationship known as the Kramers escape rate. We will briefly address this for a better understanding here.
\begin{figure}[h!]
 \begin{center}
 \includegraphics[height=4.5cm,width=6.5cm]{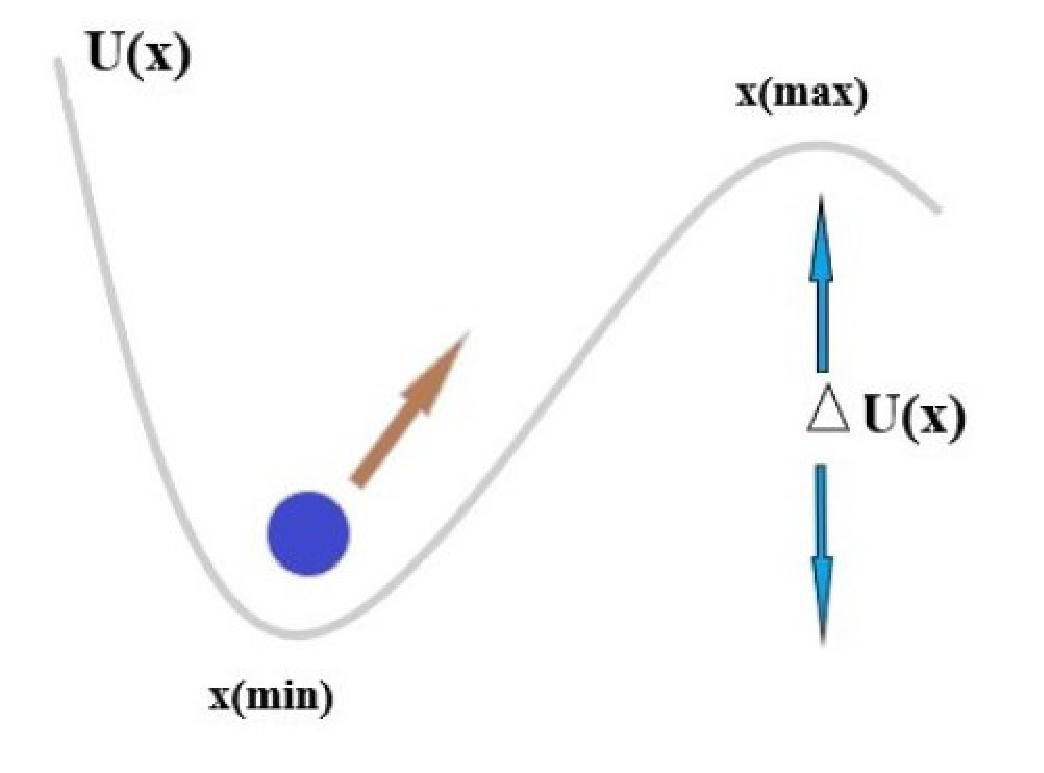}
 \caption{\small{The possibility of leakage due to Brownian motion or thermal fluctuation }}
 \label{m1}
\end{center}
\end{figure}
As we can see in Fig. (\ref{m1}), the point $ x_{min} $ is a local minimum such that particles are in equilibrium at this point and maintain their original position under small perturbations. Without loss of generality and for simplicity, we can point out a few things here. For instance, we assume only one direction of escape, which may occur through the nearest local maximum located at $ x_{max} $. On the left side of $ x_{min} $, the potential is such that it does not have any additional extremes. After the local maximum at $ x_{max} $, on the right side, there may be another local or global minimum, or even the potential may be unbounded below, which is irrelevant. This is because the escape rate should be independent of the shape of the potential on the right side of $ x_{max} $. When we add the effect of thermal baths to the system, the particles are affected by the two competing effects of fluctuation and dissipation. Random forces arising from thermal fluctuations push the particle away from its initial position and allow it to eventually climb the potential barrier. On the other hand, damping tends to slow the particle down and makes it more difficult to return to the equilibrium point. There are three possible situations here. Very Low Damping: In this scenario, particles experience minimal frictional forces. The escape rate is determined by the energy diffusion regime, where the particle's inertia plays a significant role. The particle can oscillate many times near the potential well's boundary before escaping, and the escape rate is relatively low due to the lack of significant damping forces. Intermediate Damping:  The escape rate in this regime is more complex to describe, as it involves a balance between inertial effects and damping forces. The system oscillates less than before, but it can still exist, and the escape rate is still not significant. Over-damped: In the over-damped case, the damping forces are strong, and the system experiences high frictional forces, which dominate over inertial effects. The particle moves slowly away from the potential well, and the escape rate is governed by the rate at which the system can overcome the barrier height compared to the thermal energy. The escape process is akin to a slow, diffusive motion over the barrier, and the escape rate is higher than in the very low or  Intermediate damping case and is significant. Since we intend to simulate this model according to the phase transition behavior of the black hole, then among the above options, the over-damping mode seems to be the appropriate option, since due to the strengthening of the combined effect in this mode, the system is no longer stable and the probability of the particle exiting the well will start to grow. That is, after a sufficient period of time, we can expect the particle to have almost passed through the barrier.
The escape rate is closely related to the inverse of the average time required for the first passage of the local maximum potential, which is called the "mean first passage time", which was noted in articles \cite{23,24,25,26,27,28,29,30,31,32,33,34,35,36,37,38,39,40,41,42,43,44}. It should be noted that the escape rate is different from the diffusion rate to the next minimum  \cite{53,54,55,56,57,58}. Let us consider the Fokker-Planck equation  \cite{55,58,59},
\begin{equation*}\label{1}
\frac{\partial}{\partial t}P \! \left(x ,t \right)=-\frac{\partial}{\partial x}J \! \left(x ,t \right),
\end{equation*}
where the $ P(x,t) $ is the probability distribution of particles and  current J is defined as,
\begin{equation*}\label{2}
J =-\frac{\left(\frac{\partial}{\partial x}U \! \left(x ,t \right)\right) P \! \left(x ,t \right)}{\gamma}-\mathrm{D} \left(\frac{\partial}{\partial x}P \! \left(x ,t \right)\right),
\end{equation*}
here $ U(x) $ is a potential field, $\gamma$ is the friction coefficient and D is the diffusion coefficient. The current can be rewritten as follows,
\begin{equation*}\label{3}
J =-\mathrm{D} {\mathrm e}^{-\frac{U}{k_{B} T}} \frac{\partial}{\partial x}({\mathrm e}^{-\frac{U}{k_{B} T}} P ),
\end{equation*}
$ k_B $ is the Boltzmann constant and T is the temperature. To integrate, we rewrite the above relation as follows,
\begin{equation*}\label{4}
\frac{\partial}{\partial x}({\mathrm e}^{-\frac{U}{k_{B} T}} P ) =-\frac{J {\mathrm e}^{\frac{U}{k_{B} T}}}{\mathrm{D}}.
\end{equation*}
By integrating from the above equation and taking into account the fact that on the left side of the equation, the contribution of the distribution term P in $ x_{min} $ dominates over $ x_{max} $, we can conclude,
\begin{equation*}\label{5}
J =\frac{\mathrm{D} {\mathrm e}^{\frac{U \left(x_{\min}\right)}{k_{B} T}} P \! \left(x_{\min}\right)}{\int_{x_{\min}}^{x}{\mathrm e}^{-\frac{U \left(x \right)}{k_{B} T}}d x}.
\end{equation*}
Now, if we consider the upper limit of the integral of the denominator in the above equation at a point on the right side of $ x_{max} $ and using the expansion around $ x_{max} $ and after some calculations, we will have,
\begin{equation*}\label{6}
\int_{x_{\min}}^{x}{\mathrm e}^{-\frac{U \left(x \right)}{k_{B} T}}d x =\ \sqrt{\frac{2\pi  k_{B} T}{{| \frac{d^{2}}{d x^{2}}U( x_{max})|}}}\, {\mathrm e}^{\frac{U \left(x_{\max}\right)}{k_{B} T}}.
\end{equation*}
If we define p as the probability of the particle being inside the well, then with some calculations we can find that,
\begin{equation*}\label{7}
p =P \! \left(x_{\min}\right) \ \sqrt{\frac{2\pi  k_{B} T}{{| \frac{d^{2}}{d x^{2}}U( x_{min})| \! \left(x \right)|}}}.
\end{equation*}
Finally, using the definition of Kramers escape rate($ r_k=J/P $), we will have,
\begin{equation*}\label{8}
r_{k}=\frac{\mathrm{D} \sqrt{{| \frac{d^{2}}{d x^{2}}U( x_{max}) \! } { \frac{d^{2}}{d x^{2}}U( x_{min}) \! |}}\, {\mathrm e}^{-\frac{U \left(x_{\max}\right)-U \left(x_{\min}\right)}{k_{B} T}}}{2 \pi  k_{B} T}.
\end{equation*}
When the system reaches thermal equilibrium, D can be considered constant[55], and with a little simplification, the above relationship can be rewritten as follows[56],
\begin{equation}\label{9}
r_{k}=\frac{ \sqrt{{| \frac{d^{2}}{d x^{2}}U( x_{max}) \! } { \frac{d^{2}}{d x^{2}}U( x_{min}) \! |}}\, {\mathrm e}^{-\frac{U \left(x_{\max}\right)-U \left(x_{\min}\right)}{D}}}{ 2 \pi}.
\end{equation}
Note that the formula can only be applied for $ \Delta U >> k_B T$.
\subsection{Different shapes for free energy landscape }
Free energy landscape provides a way to visualize and understand how systems, transition between different states over time. In simple terms, the free energy landscape is a graphical representation where each point on the landscape corresponds to a possible state of the system. From the vantage point of free energy, the exploration of phase transition conditions becomes accessible through the scrutiny of the emergence and stability of distinct phases. Furthermore, the characterization and analysis of the free energy landscape are profoundly influenced by its relationship with the order parameters. Taking the radius of a black hole's event horizon as an exemplary order parameter, one can articulate the free energy landscape. Consequently, the phase transition of a black hole can be interpreted as a stochastic thermal fluctuation of said order parameter, offering a novel perspective on the underlying physics \cite{21}. In the following, since we want to study the phase transition from the perspective of the free energy landscape and in a different structure under specific thermal fluctuations, we will introduce different forms of this energy that can help us in this way.
\subsubsection{Gibbs Free Energy Landscape }
The Gibbs free energy function, renowned for its intrinsic reliance on enthalpy and temperature, emerges as the special tool for probing the phase transitions within a black hole's architecture, particularly under constant pressure conditions. The swallowtail behavior, manifested graphically as a function of temperature, stands as a prevalent methodology for the exploration of phase transitions. Within the formalism of the free energy perspective, the Gibbs free energy is delineated as a continuous function that is intimately connected to the system's order parameter, thereby providing a robust framework for the analysis of phase behavior,
\begin{equation}\label{10}
G =H-T_H S ,
\end{equation}
which $T_H$ is Hawking temperature, H is enthalpy and S is entropy. Also considering the correspondence between mass and enthalpy in the extended phase of a black hole ($H \equiv M $), this equation can be rewritten in terms of mass.\\
During the intricate process of phase transition, particularly of the first order, an assortment of black holes of varying sizes-small, medium, and large-along with transitional states, frequently emerges. Consequently, it is feasible to conceptualize a collection of these black hole space-times, each with an arbitrary horizon radius, at a specified temperature "T".
For a better understanding, it must be stated that one of the most important characteristics of a black hole is its temperature, or more specifically, the Hawking temperature. The mere existence of temperature, regardless of how it is defined, implies the possibility of thermal fluctuations within the system. Consider this as a background. Now, the black hole under study has been excited for some environmental reason and begins a phase transition process. At both the initial and ends of this phase transition, there are two locally stable black holes, one of which is globally more stable than the other, setting the stage for this transition. However, an important point regarding these two states is that both must necessarily be solutions to Einstein's field equations. The golden key to defining a set of black holes lies here. Although the system must end up in a state that is necessarily a solution to Einstein's equations at both ends, there is no need to strictly follow the static field equations during the transition from the initial phase to the final phase, since the system is in an unstable transitional phase. Therefore, considering a predefined energy range and with the condition that the chosen paths do not destroy the overall stability of the system, the system can be allowed to traverse any desired path from the origin to the end.
This freedom of choice and multiple paths could serve as a gateway for the introduction of probability and statistical methods. Hence, by selecting an arbitrary indistinctive set within this framework, which acts as a canonical ensemble for us, a statistical framework for analysis can be presented. Notably, the temperature "T" of the ensemble may diverge from the Hawking temperature associated with black holes. Each of the space-time configurations within this ensemble is characterized by a distinct Gibbs free energy. Hence, we refer to this construct as the generalized Gibbs result or, more descriptively, the Gibbs free energy landscape.
\begin{equation}\label{11}
G_L =M-T S .
\end{equation}
The terminology 'on-shell' and 'off-shell' delineates the aforementioned temperature discrepancy. 'Off-shell' encompasses all transitions occurring at temperature "T", which, despite conforming to the laws of thermodynamics, do not necessarily constitute solutions to Einstein's field equations \cite{21}. It is imperative to recognize that solely the extremal points on the $ G_L$ signify the authentic black hole phases that comply with the Einstein field equations. Furthermore, within this context, the local maxima and minima are indicative of the unstable and locally stable black hole phases, respectively \cite{60}.
\subsubsection{Landau free energy }
The Landau free energy constitutes another showy path for scrutinizing the phase transition dynamics within  the system’s free energy landscape. While using this method assumes heightened significance in the context of second-order phase transitions- wherein the bifurcation of the Landau functional's local minimum epitomizes the thermodynamic system's second-order phase transition- it is also pertinent to first-order phase transitions. Specifically, the transmutation of the global minimum is emblematic of a first-order phase transition, underscoring the multifaceted nature of the Landau-free energy in characterizing phase behavior. This function can be written as follows \cite{45},
\begin{equation}\label{12}
L =\int F \! \left(X ,P ,T \right)d X,
\end{equation}
that in the delineated relationship, P denotes pressure, T represents temperature, and X serves as an auxiliary variable. However, in the realm of black hole thermodynamics, pressure holds a distinctive advantage. Contrary to other thermodynamic variables such as temperature, mass, and volume -which are solely functions of the event horizon's radius- pressure not only contributes to the equation of state, inextricably linked to the radius, but also assumes an independent role based on the definition of the cosmic constant in the studied model. This endows it with a unique identity within the thermodynamic landscape. Consequently, preserving the essence of generality, the "F" function can be reformulated as follows,
\begin{equation}\label{13}
F \! \left(X ,P ,T \right) = P - f \! \left(X ,T \right).
\end{equation}
Given that "L" is an energy function, it inherently embodies the fundamental characteristic of energy minimization. This implies that the system invariably seeks a trajectory that reduces the energy expenditure required for transformation. Therefore, among the triad of system parameters $ (X, P, T) $ that influence the "F" function, only the configurations that minimize "L" will accurately reflect the system's state,
\begin{equation}\label{14}
\frac{d}{d X}L \ = F \! \left(X ,P ,T \right)=0 .
\end{equation}
The important point before the end of this section is that the most familiar form of the characteristic "X" in Landau's definition is the volume, to be able to use the probabilistic method for this function, we must consider the volume off-shell
\subsubsection{ Thermal potential}
Another form of landscape energy that can be used to investigate phase transition based on thermal fluctuation is the model that was introduced in \cite{47} under the title of thermal potential.
\begin{equation}\label{15}
U =\int \left(T_{H}\ - T \ \right)d S,
\end{equation}
which $T_H$ is Hawking temperature,$T$ is 'Off-shell' temperature, and $S$  is entropy that has appeared here in the role of a variable. Since In thermodynamics, the vigor of thermal motion is quantified by the product of temperature and entropy. Consequently, the aforementioned thermal potential serves as an approximate gauge for the extent to which all conceivable states within the canonical ensemble deviate from the genuine black hole state.\\
When a black hole system is situated within a potential field $U$, it is subject to thermodynamic fluctuations. The resultant stochastic behavior of black holes within such a potential field can offer insights into certain thermodynamic attributes of black holes. Furthermore, the thermal potential's utility extends to its ability to indicate system stability through the analysis of concavity and convexity at critical points.
\section{Non-linear magnetic-charged AdS black hole surrounded\\ by quintessence in the perfect fluid dark matter  background}
In this article, we went to black holes that somehow interact with the structure of matter and dark energy, so that we can study the effect of the presence of parameters caused by the dark structure on the first-order phase transition. So the metric of (NLM-C-Q-PFD) black holeis as follows \cite{51},
\begin{equation*}\label{(2)}
f \! \left(r \right)=1-\frac{2 M \,r^{2}}{Q^{3}+r^{3}}+\frac{\alpha  \ln \! \left(\frac{r}{{| \alpha |}}\right)}{r}-\frac{c_{q}}{r^{3 \varepsilon +1}}+\frac{8 \pi  P \,r^{2}}{3}.
\end{equation*}
For the main quantities of this model i.e. mass $M$, Hawking temperature $T_H$, entropy $S$, pressure $P$ and volume $V$ we will have (note that as long as there is no specific definition of the radius, it will be $ r=r_H $ in calculations) \cite{51},
\begin{equation}\label{16}
M =\frac{\left(Q^{3}+r^{3}\right) \left(1+\frac{\alpha  \ln \left(\frac{r}{|\alpha|}\right)}{r}-\frac{c_{q}}{r^{3 \varepsilon +1}}+\frac{8 \pi  P \,r^{2}}{3}\right)}{2 r^{2}},
\end{equation}
\begin{equation}\label{17}
T_H =\frac{\frac{-2 Q^{3}+r^{3}}{r}+\frac{3 c_{q} \varepsilon  \left(r^{3}+\frac{Q^{3} \left(\varepsilon +1\right)}{\varepsilon}\right)}{r^{3 \varepsilon +2}}+\frac{\alpha  Q^{3} \left(1-3 \ln \left(\frac{r}{|\alpha|}\right)\right)}{r^{2}}+\alpha  r +8 \pi  P \,r^{4}}{4 \pi  \left(Q^{3}+r^{3}\right)},
\end{equation}
\begin{equation}\label{18}
S =\pi  r^{2} \left(1-\frac{2 Q^{3}}{r^{3}}\right),
\end{equation}
\begin{equation}\label{19}
P =\frac{3 \alpha  Q^{3} \ln \! \left(\frac{r}{|\alpha|}\right)-3 r^{3-3 \varepsilon} c_{q} \varepsilon -3 Q^{3} c_{q} \left(\varepsilon +1\right) r^{-3 \varepsilon}+\left(4 \pi  T \,r^{2}-\alpha +2 r \right) Q^{3}+4 T \pi  r^{5}-r^{4}-\alpha  r^{3}}{8 \pi  r^{6}},
\end{equation}
\begin{equation}\label{20}
V =\frac{4 \pi  \left(Q^{3}+r^{3}\right)}{3},
\end{equation}
where $\alpha$ denotes the intensity of the PFDM, $c_q$ and $\varepsilon$ are the quintessence parameters and Q is the magnetic charge.
\subsubsection{Behavioral comparison of landscape free energy}
\begin{figure}[h!]
 \begin{center}
 \subfigure[]{
 \includegraphics[height=4.5cm,width=5.5cm]{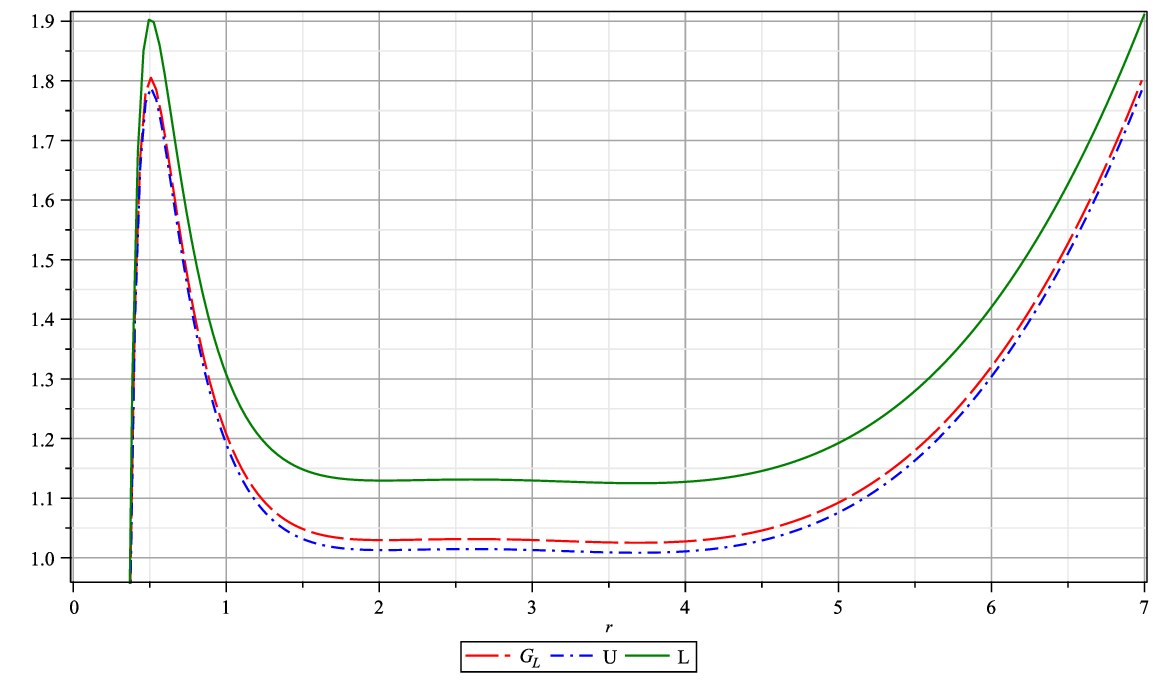}
 \label{2a}}
 \subfigure[]{
 \includegraphics[height=4.5cm,width=5.5cm]{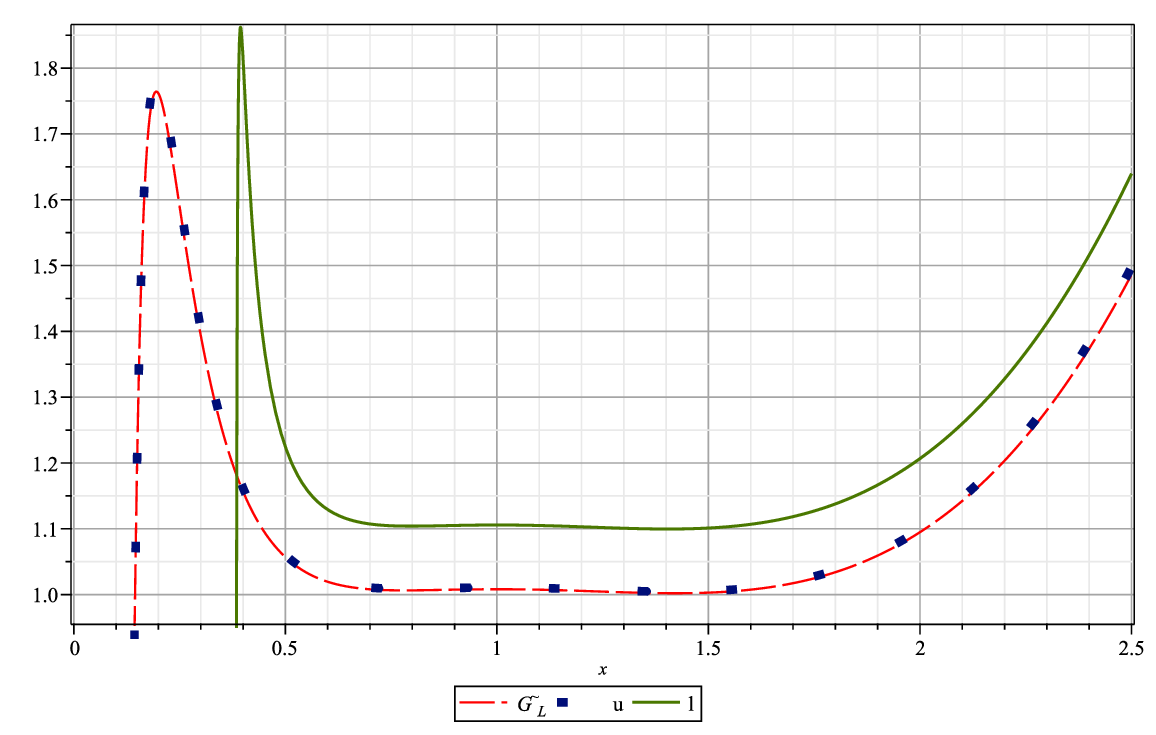}
 \label{2b}}
   \caption{\small{ Comparison of the Gibbs landscape free energy ($G_L$), thermal potential ($U$), and Landau free energy ($L$) in the diagram in two states, i.e., with dimension in Fig. (\ref{2a}) and dimensionless in Fig. (\ref{2b})}}
 \label{m2}
\end{center}
 \end{figure}

\begin{figure}[H]
 \begin{center}
 \subfigure[]{
 \includegraphics[height=4.5cm,width=5.5cm]{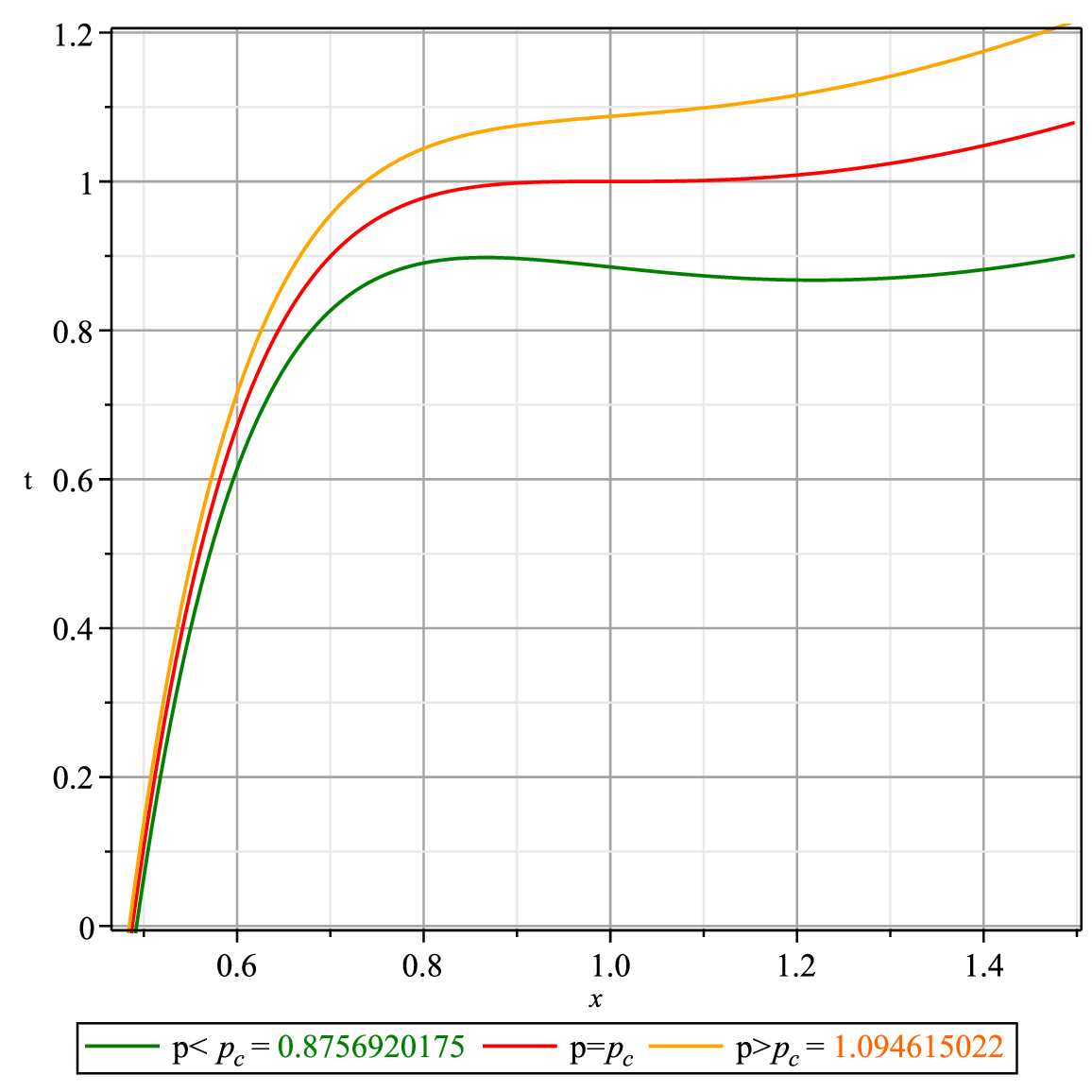}
 \label{3a}}
 \subfigure[]{
 \includegraphics[height=4.5cm,width=5.5cm]{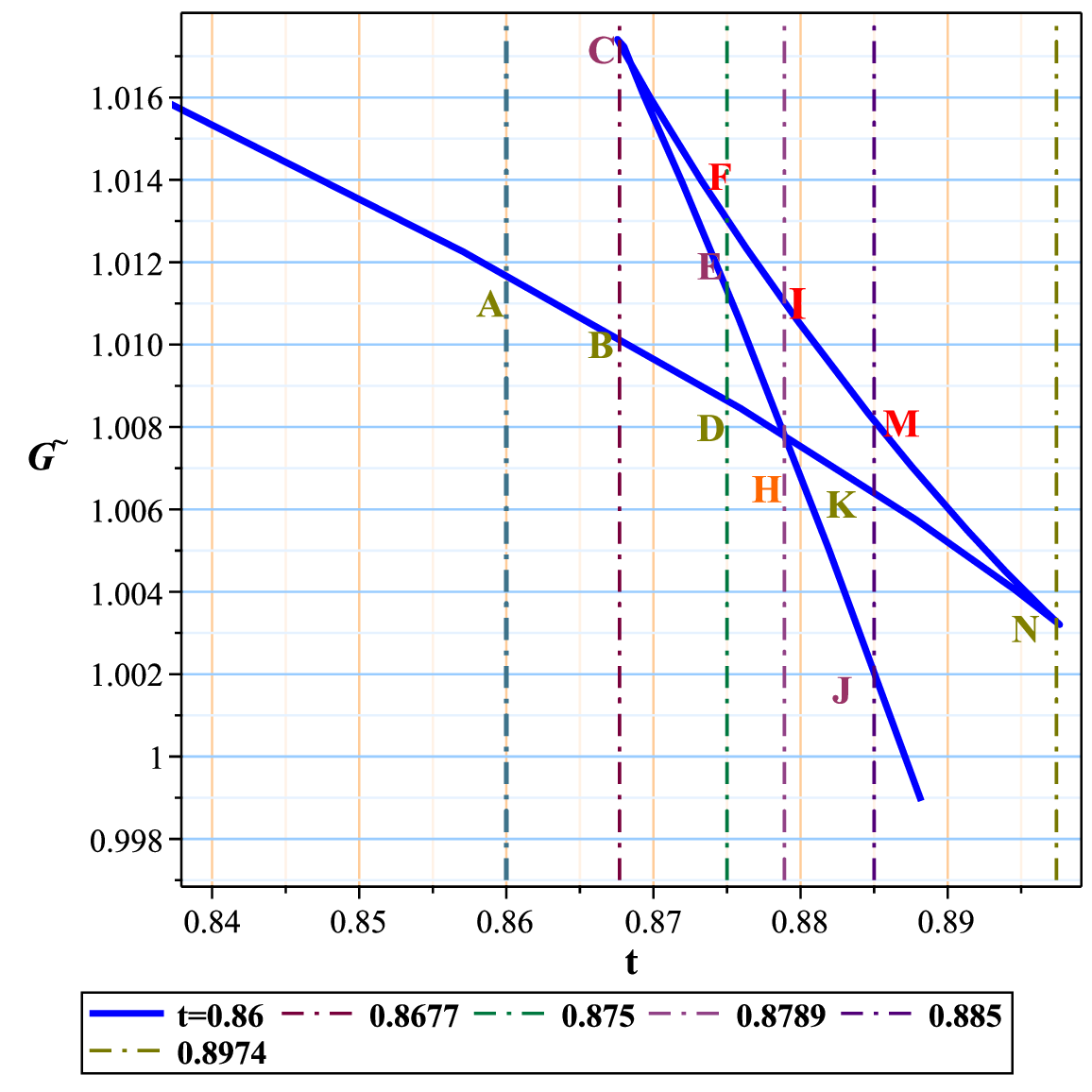}
 \label{3b}}
   \caption{\small{The graph of T against x for free parameters in Fig(3a) and Gibbs Free energy versus temperature in Fig(3b)}}
 \label{m3}
\end{center}
 \end{figure}

\begin{figure}[H]
 \begin{center}
 \subfigure[]{
 \includegraphics[height=4.15cm,width=4cm]{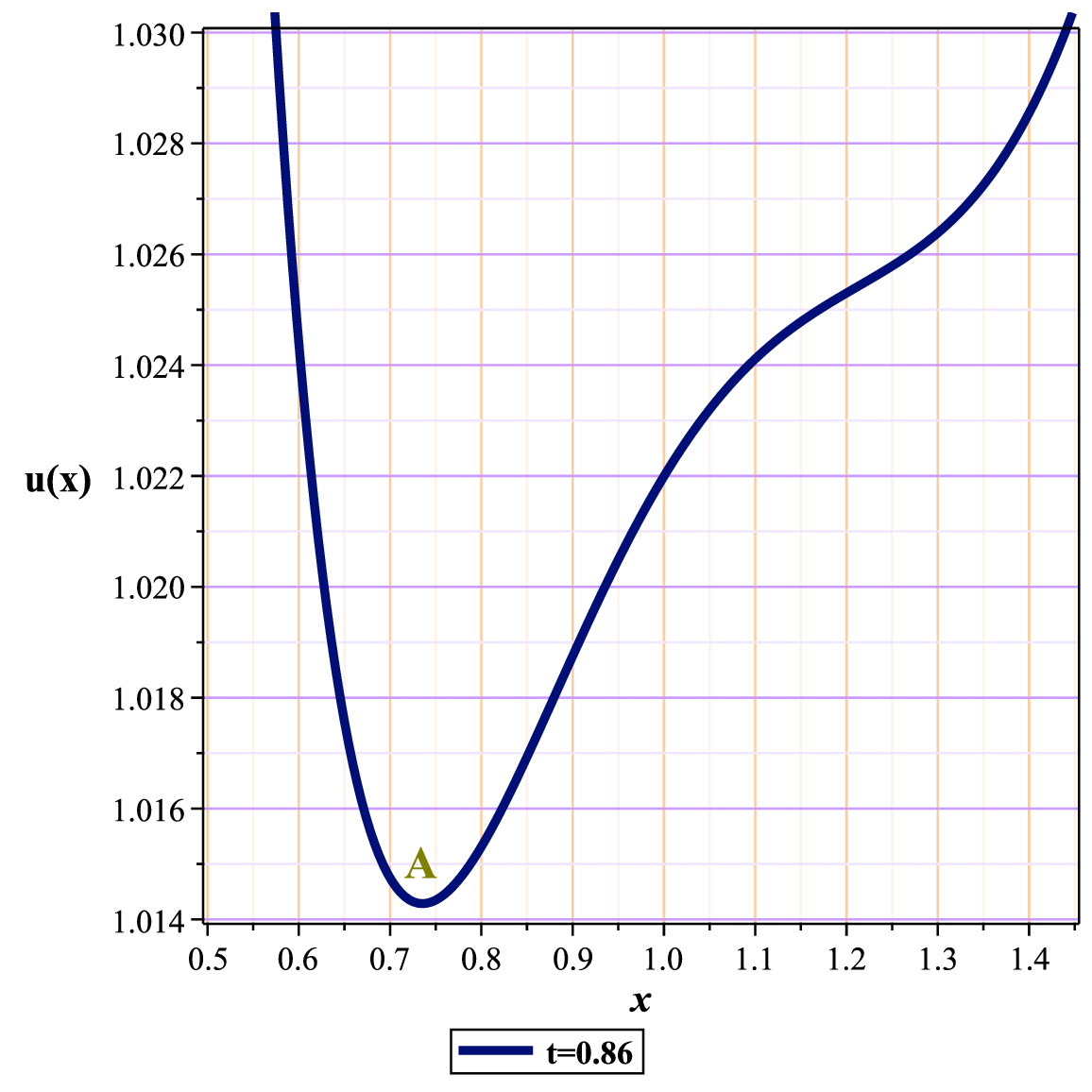}
 \label{4a}}
 \subfigure[]{
 \includegraphics[height=4.15cm,width=4cm]{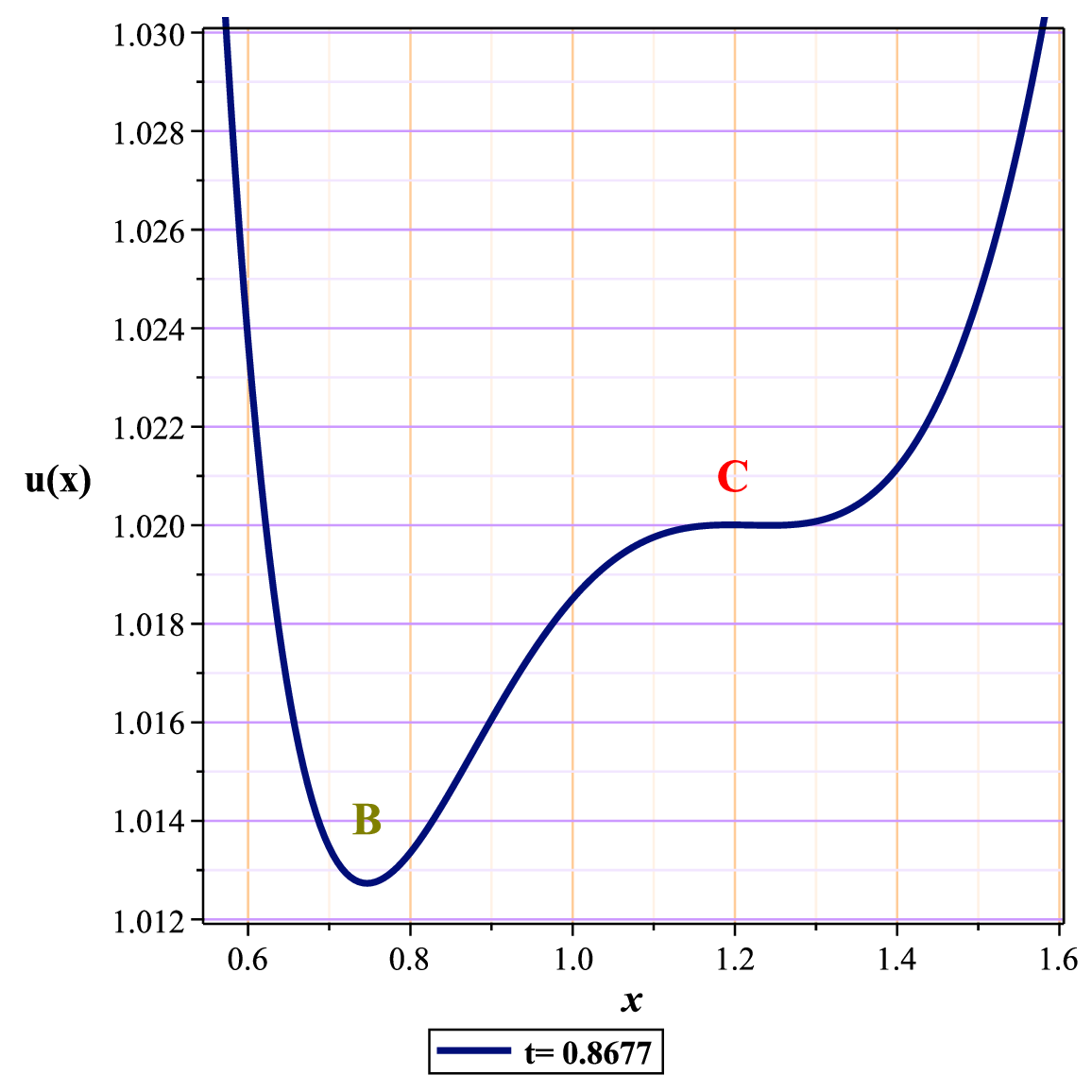}
 \label{4b}}
 \subfigure[]{
 \includegraphics[height=4.20cm,width=4cm]{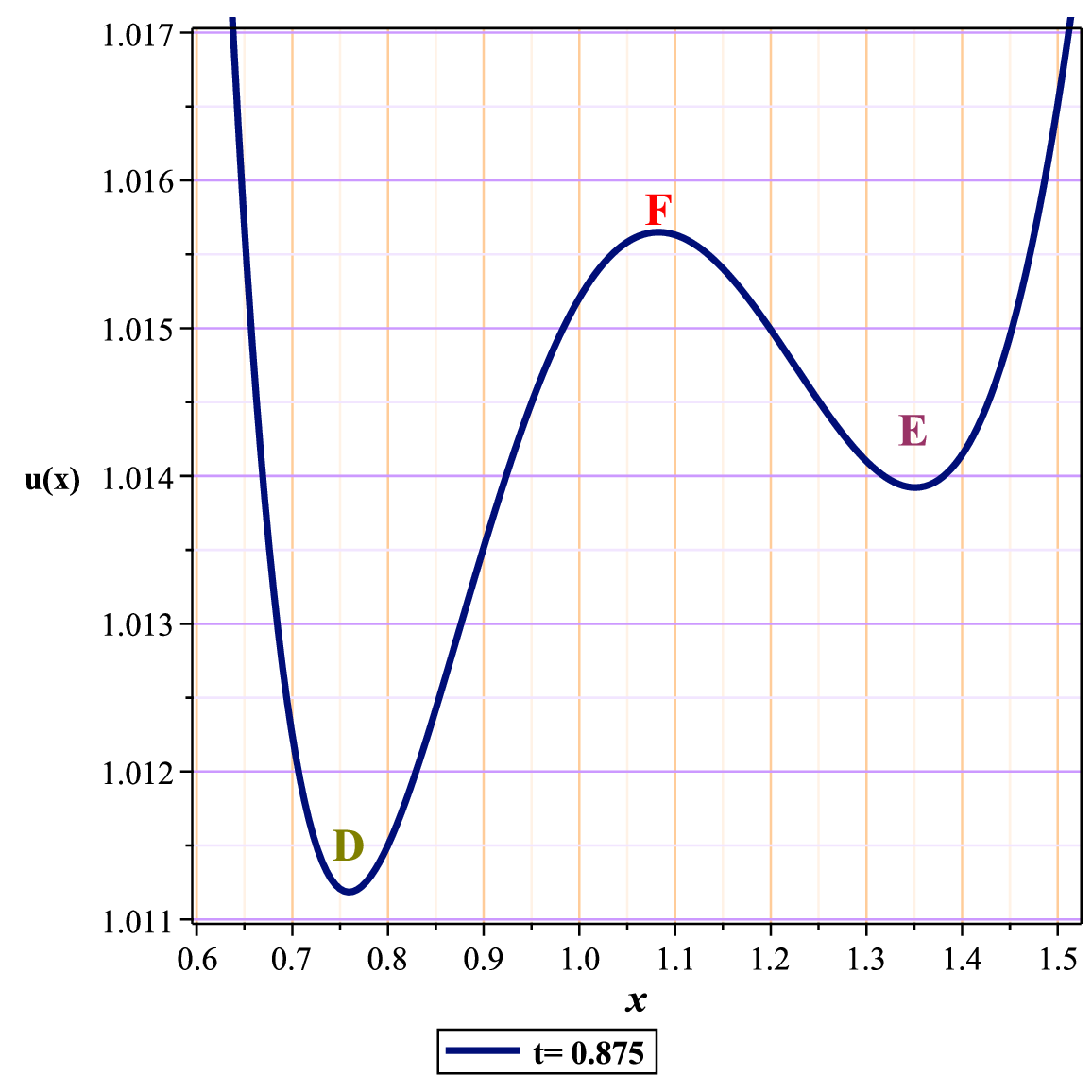}
 \label{4c}}
 \subfigure[]{
 \includegraphics[height=4.20cm,width=4cm]{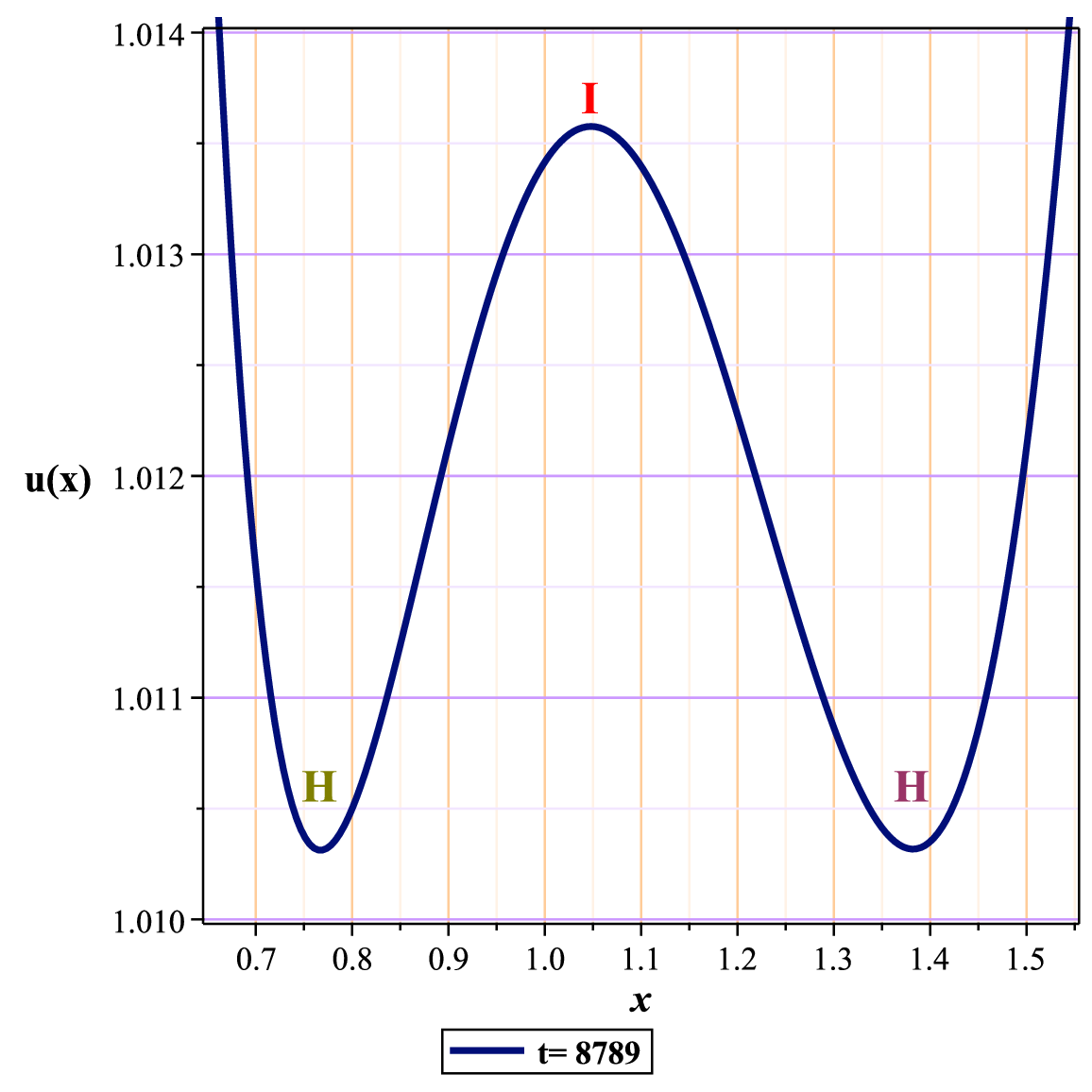}
 \label{4d}}
 \subfigure[]{
 \includegraphics[height=4.20cm,width=4cm]{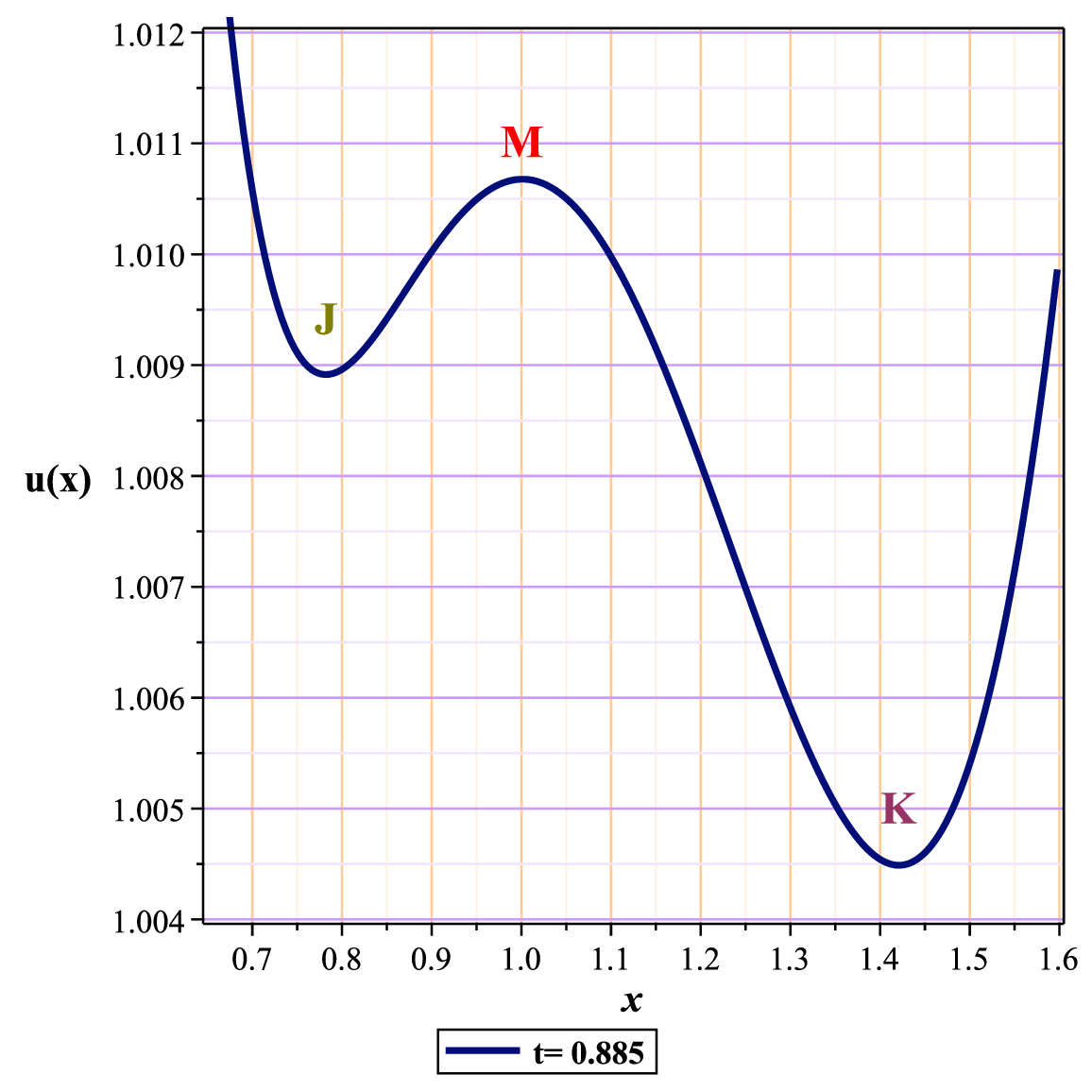}
 \label{4e}}
 \subfigure[]{
 \includegraphics[height=4.20cm,width=4cm]{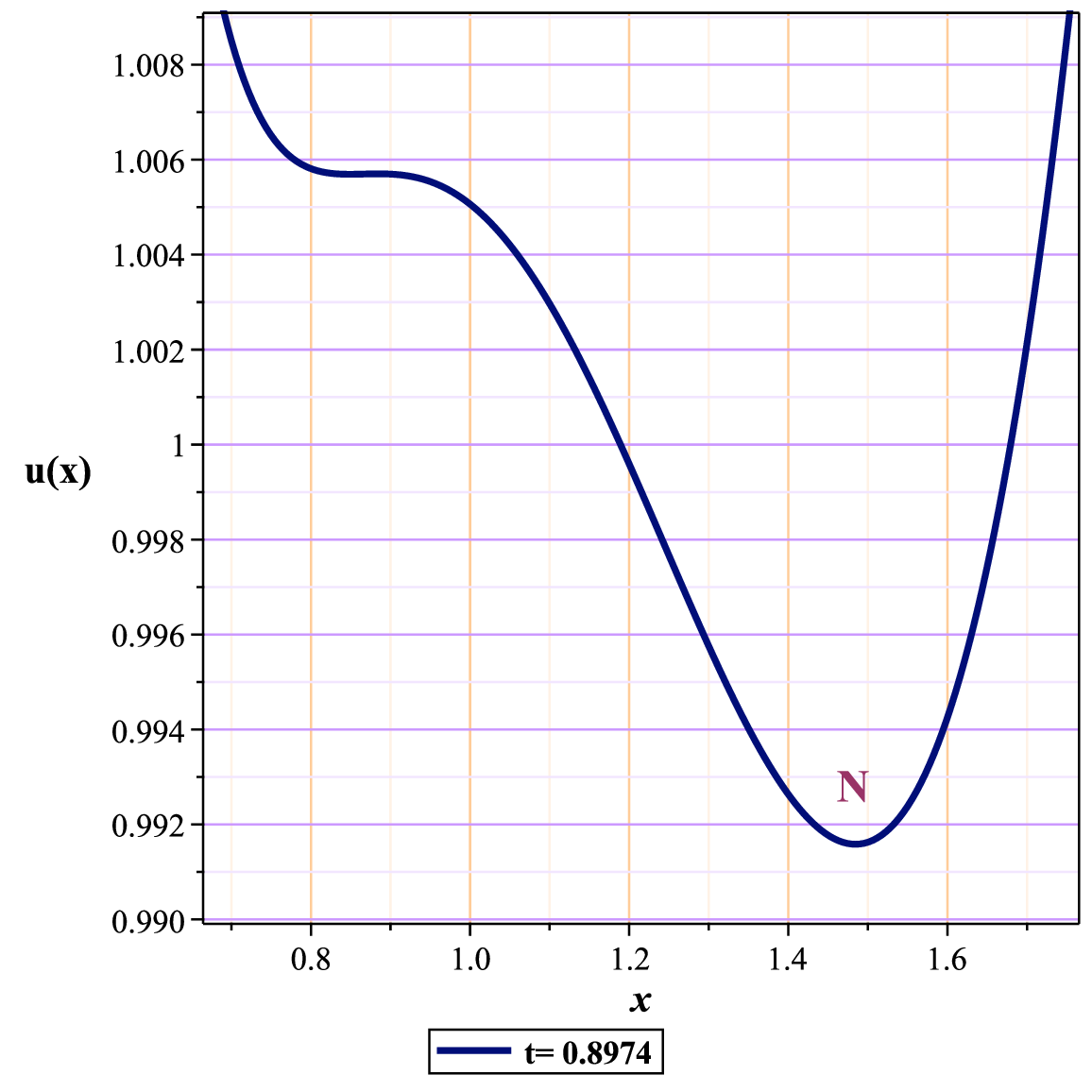}
 \label{4f}}
 \subfigure[]{
 \includegraphics[height=4.15cm,width=4cm]{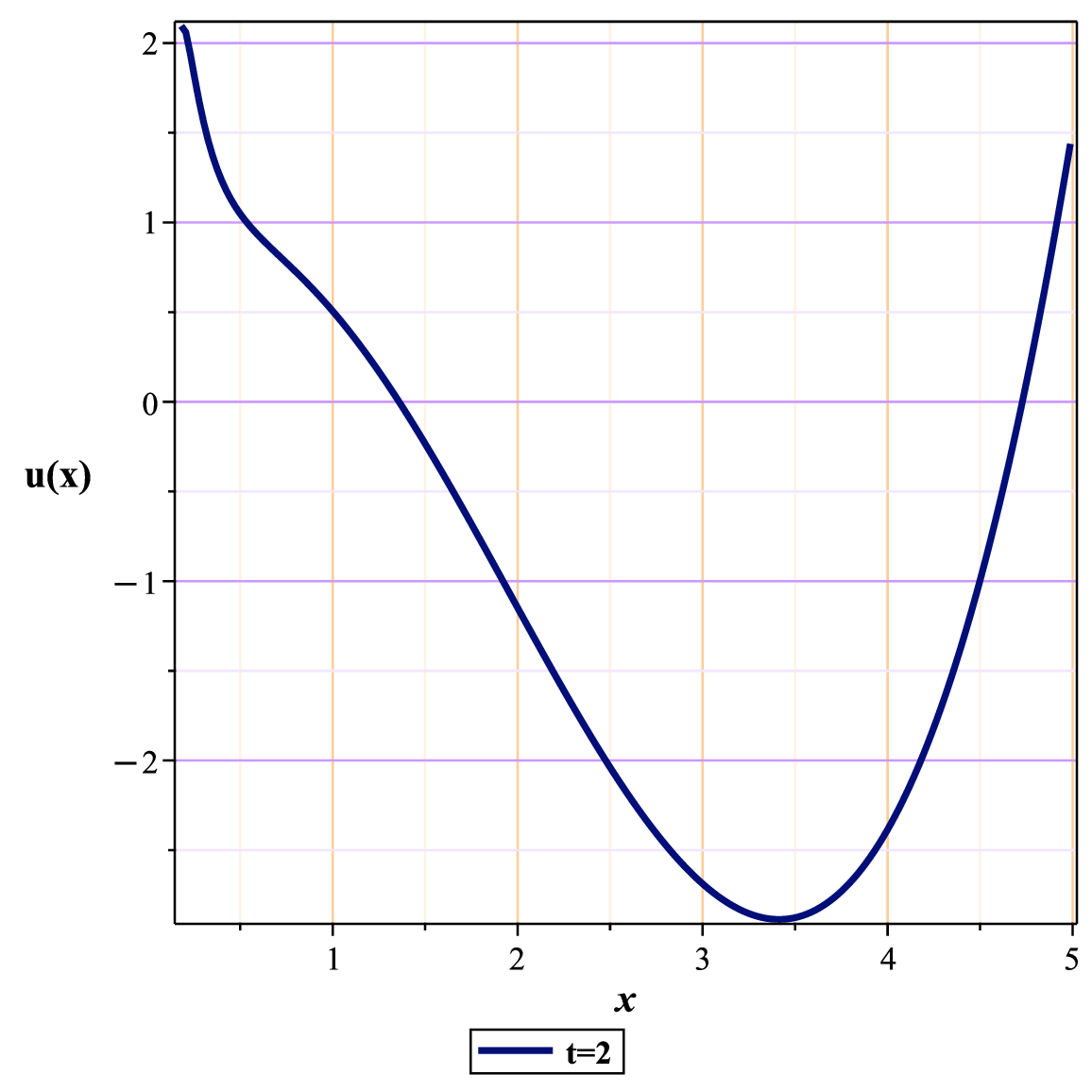}
 \label{4g}}
 \caption{\small{Behavioral sequence of energy in terms of x for different temperatures}}
 \label{m4}
\end{center}
 \end{figure}

Now according to Eqs. (\ref{2}), (\ref{3}), (\ref{4}) and (\ref{7}) for Gibbs and Landau free energy function and also the thermal potential of this model, we will have\footnote{Note that in the following relationships, due to the number of parameters, the beginning of the equations is shown with a bold symbol only for better clarity.},
\begin{equation}\label{21}
\begin{split}
&\textbf{G} =\mathcal{A}+\mathcal{B},\\
&\mathcal{A}=\frac{\left(Q^{3}+r^{3}\right) \left(3 r +3 \alpha  \ln \! \left(\frac{r}{|\alpha|}\right)-3 c_{q} r^{-1-3 \varepsilon} r +8 \pi  P \,r^{3}\right)}{6 r^{3}}\\
&\mathcal{B}=\frac{\left(-\frac{2 Q^{3}-r^{3}}{r}+3 c_{q} \left(Q^{3} \varepsilon +r^{3} \varepsilon +Q^{3}\right) r^{-2-3 \varepsilon}+\frac{\alpha  Q^{3} \left(1-3 \ln \left(\frac{r}{|\alpha|}\right)\right)}{r^{2}}+\alpha  r +8 \pi  P \,r^{4}\right) \left(2 Q^{3}-r^{3}\right)}{4 r \left(Q^{3}+r^{3}\right)}\\
\end{split}
\end{equation}
\begin{equation}\label{22}
{G}_{L} =\frac{\left(Q^{3}+r^{3}\right) \left(1+\frac{\alpha  \ln \left(\frac{r}{{| \alpha |}}\right)}{r}-\frac{c_{q}}{r^{1+3 \varepsilon}}+\frac{8 \pi  P \,r^{2}}{3}\right)}{2 r^{2}}-T \pi  r^{2} \left(1-\frac{2 Q^{3}}{r^{3}}\right)
\end{equation}
\begin{equation}\label{23}
\begin{split}
&\textbf{L} = P X -\frac{\mathcal{C} +\mathrm{D}}{48 Q^{3} \pi^{2}-36 \pi  X},\\
&\mathcal{C}=36 \left(Q^{3} \pi^{\frac{7}{3}}-\frac{\pi^{\frac{4}{3}} X}{4}\right) 2^{\frac{2}{3}} T \left(-4 \pi  Q^{3}+3 X \right)^{\frac{2}{3}}+9 \,2^{\frac{1}{3}} X \pi^{\frac{2}{3}} \left(-4 \pi  Q^{3}+3 X \right)^{\frac{1}{3}},\\
&\mathrm{D}=6 \ln \! \left(-4 \pi  Q^{3}+3 X \right) \pi  X \alpha -18 X \pi^{\varepsilon +1} c_{q} \left(-\pi  Q^{3}+\frac{3 X}{4}\right)^{-\varepsilon}-8 Q^{3} \pi^{2} \alpha  \left(\ln \! \left(\pi \right)+3 \ln \! \left({| \alpha |}\right)+2 \ln \! \left(2\right)\right),\\
\end{split}
\end{equation}
\begin{equation}\label{24}
\begin{split}
&\textbf{U} =\mathcal{E}-T \pi  r^{2} \left(1-\frac{2 Q^{3}}{r^{3}}\right),\\
&\mathcal{E}=\frac{3 c_{q} \left(-Q^{3}-r^{3}\right) r^{-3 \varepsilon}+3 \alpha  \left(Q^{3}+r^{3}\right) \ln \! \left(r \right)+8 \pi  P \,r^{6}-3 \alpha  Q^{3} \ln \! \left(|\alpha| \right)+3 Q^{3} r +3 r^{4}}{6 r^{3}}\\.
\end{split}
\end{equation}
Now we will go to a special state that is formed based on the arbitrary parameter setting. For this purpose, by considering the values  $\alpha=0.6, c_q=0.2, \varepsilon=-2/3, Q=1$ for the parameters and with a little calculation for the critical quantities, we will have,
\begin{equation}\label{25}
\begin{split}
&T_{c}=  0.02426755451,\hspace{0.7cm} P_{c} =  0.00456781599,\hspace{0.7cm} r_{c} =  2.594697206,\\
&V_{c}= 77.36141969,\hspace{0.7cm} U_{c} = 1.00399626317,\hspace{0.7cm} G_{c} =  1.02312988617.\\
\end{split}
\end{equation}
According to the usual procedure, we convert the parameters into dimensionless form so that the results can be compared with other studies in addition to convenience,
\begin{equation}\label{26}
t = \frac{T}{T_{c}},\hspace{0.7cm}p = \frac{P}{P_{c}},\hspace{0.7cm}x = \frac{r}{r_{c}},\hspace{0.7cm}\widetilde{G} = \frac{G}{G_{c}},\hspace{0.7cm}{\widetilde{G}_L} = \frac{G_L}{G_{c}},\hspace{0.7cm}l = \frac{L}{G_{c}},\hspace{0.7cm}u = \frac{U}{U_{c}}.
\end{equation}
Before any calculations, it is better first compare the various forms of free energy definitions, namely $ G_L $, U, and L. Theoretically, from the canonical perspective the number of particles and mass should remain constant, consequently only the entropy-temperature and pressure-volume terms will play a role in the first law of thermodynamics. Additionally, given that in Landau's free energy, $X$ can be considered equivalent to volume, it appears that ultimately, all three introduced forms for the free energy landscape must coincide. In Figs (\ref{2a}) and (\ref{2b}), for the model under study, we have plotted these three definitions once according to their general structure and once based on their dimensionless state. Practically, in terms of the shape structure, all three forms have the same trend, but there are also some visible differences, some of which may be attributed to computational precision. Moreover, since Landau's free energy has dimensions of energy, we used the $ G_c$  to render it dimensionless, which seems to have increased the degree of displacement, although the shape has been preserved. An intriguing point that comes to mind before concluding this discussion is that in statistical ensembles, such as the grand canonical ensemble where terms related to the chemical potential play a more prominent role, the complexities and structural differences of these three definitions may become more apparent compared to this scenario. Concerning the above discussion and as seen in Fig (\ref{m2}), all the shapes of the free energy landscape (with a little tolerance for displacement) will practically have the same shape structure in the canonical view. Accordingly, the choice and use of each do not seem to have an advantage over the others. For this reason, we will henceforth utilize them under the designation of the free energy landscape, without loss of generality.
Now, to be able to interpret the behavior of the energy function, we need to have the graph of temperature changes Fig. (\ref{3a}) and Gibbs free energy in terms of temperature Fig. (\ref{3b}).
To better interpret the behavior of free energy, we initially turn to the swallowtail Fig. (\ref{3b}). In this figure, the line (A-N) represents the small black hole, and( C-J ) represents the large black hole, both of which are locally stable with positive heat capacity, and the line (N-C) with negative heat capacity represents the intermediate black hole, which is unstable. As a classical principle, we know that systems always tend to choose states with the least amount of energy. This principle also holds for the free energy landscape, meaning the system will move towards its minimum. In other words, the minima of the energy function correspond to stable states, and its maxima represent unstable states. Considering the overall shape in Fig. (\ref{m4}), it is also important to note that in a system with multiple local minima, each of these minima, although locally stable relative to their immediate vicinity, may act as unstable states globally relative to the primary minimum. Even if the system is initially in these states, over time it will progress towards the ultimate global minimum.\\
In Fig. (\ref{4a}), as is evident, the potential at point A has a global minimum, which belongs to the branch of the small black hole. Over time, with the increase in temperature, in Fig. (\ref{4b}), the local extrema B (minimum) and C (maximum) begin to form. However, they are still in the form of globally unstable states, and the system tends to maintain its position at the primary minimum on the branch of the small black hole. The situation in Fig. (\ref{4c}) is similar to (\ref{4b}), with the difference that the local extrema have fully formed. With a gradual increase in temperature, the system progresses towards the formation of a large and stable black hole, as shown in Fig. (\ref{4d}). Yet, the small black hole also exists in a stable state, indicating a situation where two equivalent global minima are created. This state corresponds to the isotherm line (H-I) in Fig. (\ref{3b}), which seemingly has a maximum on the branch of the intermediate black hole (I) and a global minimum (H). However, in reality, (H) is located on both the small and large black holes, meaning that a state is created where both the small and large black holes appear simultaneously and stably. In other words, a coexistence phase between the two stable black holes is formed. Further, the process of temperature increase will strengthen the large black hole and weaken the small black hole, Fig. (\ref{4e}) and (\ref{4f}). The point that can be mentioned at the end of this discussion is related to the second-order phase transition. When we study the system at a temperature higher than the critical temperature, Fig. (\ref{4g}), the system will follow a path similar to (I-H-J) in Fig. (\ref{3b}), and practically it is directly and continuously transferred from the small black hole to the large one.
\subsubsection{Kramer’s escape rate}
\begin{figure}[H]
 \begin{center}
 \subfigure[]{
 \includegraphics[height=4.5cm,width=4cm]{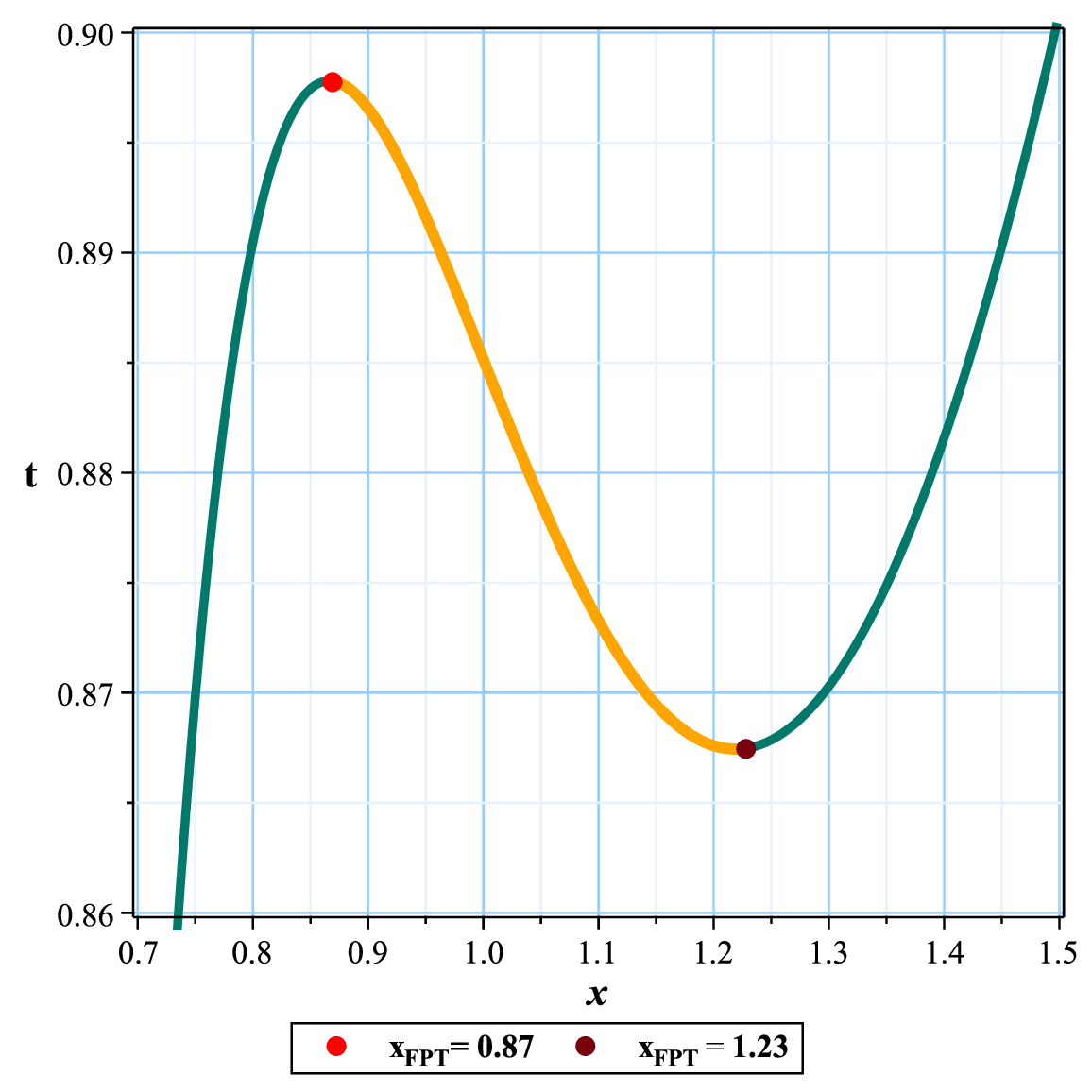}
 \label{5a}}
 \subfigure[]{
 \includegraphics[height=4.5cm,width=4cm]{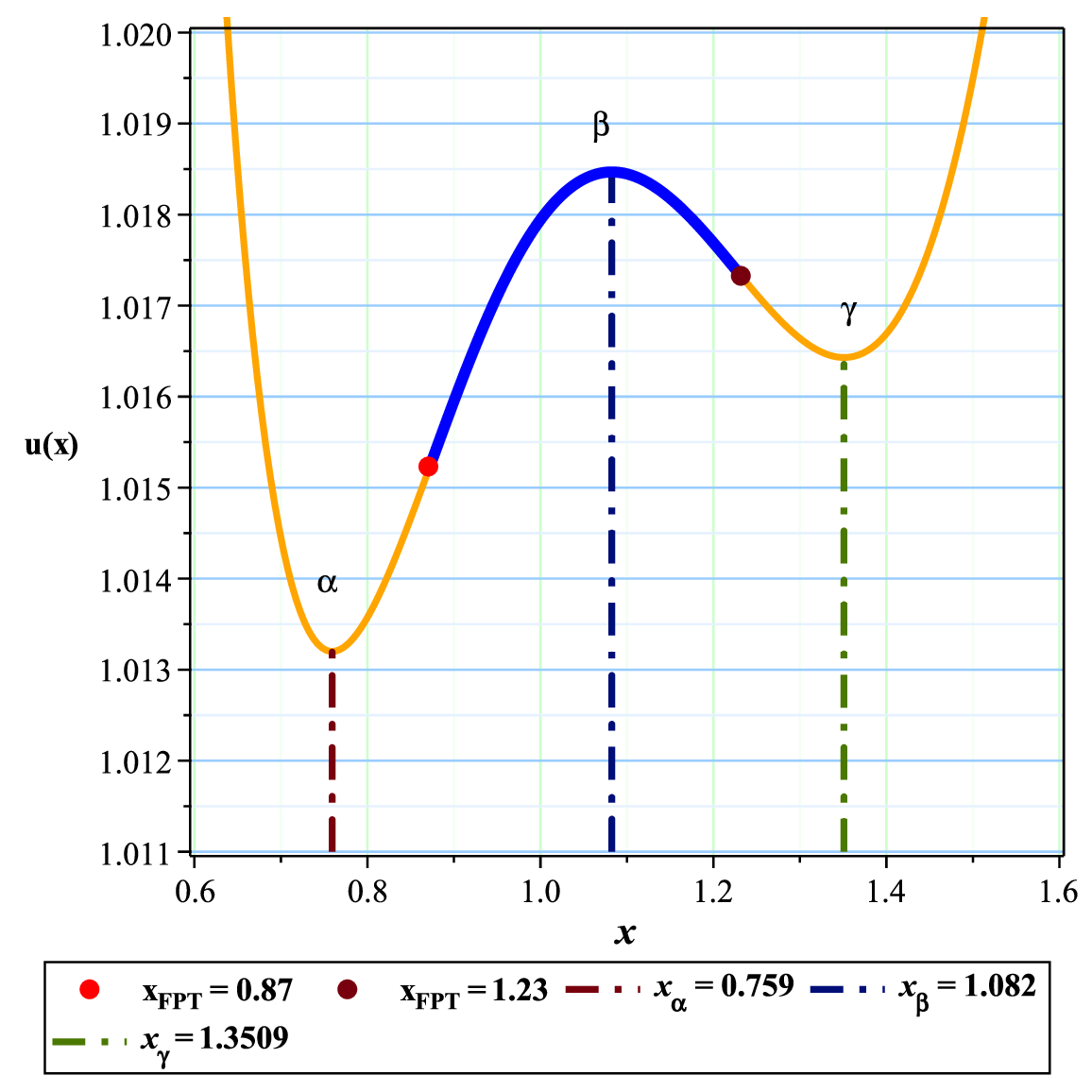}
 \label{5b}}
 \subfigure[]{
 \includegraphics[height=4.5cm,width=4cm]{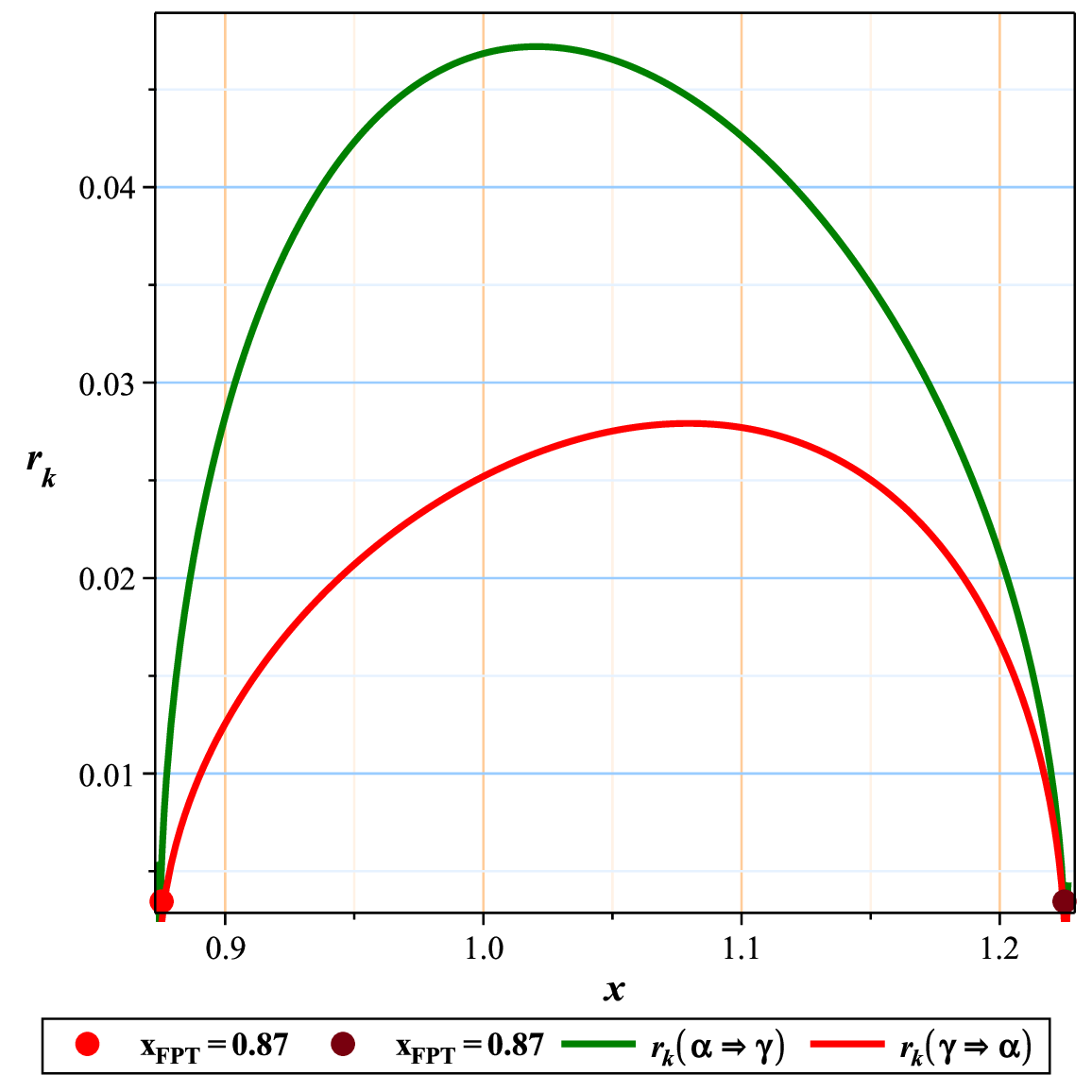}
 \label{5c}}
 \subfigure[]{
 \includegraphics[height=4.5cm,width=4cm]{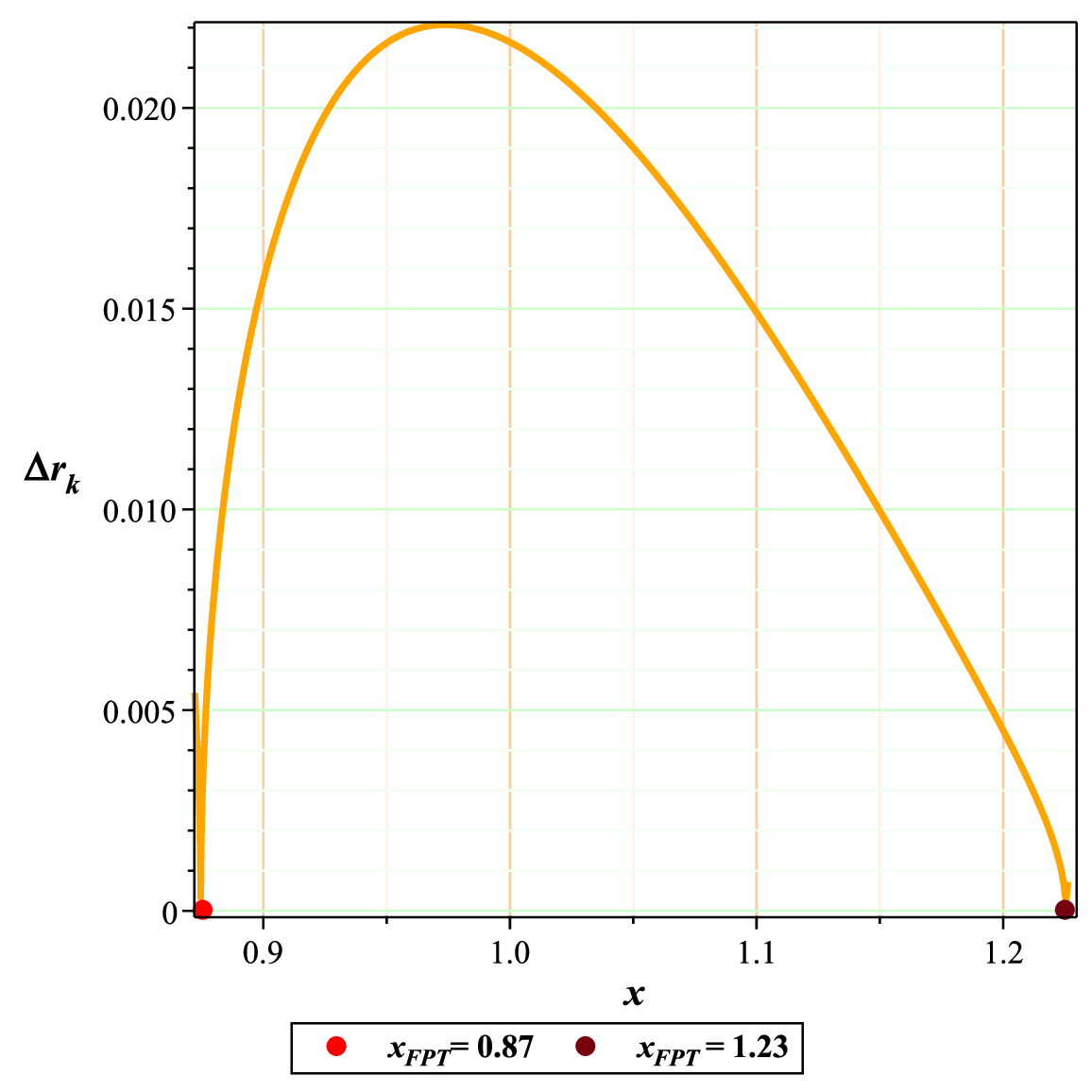}
 \label{5d}}
   \caption{\small{t versus x  and the First order Phase Transition region (FPT) in Fig. (\ref{5a}),\hspace{0.1cm} (\ref{5b}): The graph of $u(r)$ against x for free parameters and the FPT points, \hspace{0.1cm} (\ref{5c}): comparing $r_k$ with respect to $x$ for($\alpha\rightarrow\gamma$) and ($\gamma\rightarrow\alpha$) and the coordinate of contact point \hspace{0.1cm} Fig. (\ref{5d}): $\Delta r_k$ ($r_k(\alpha\rightarrow \gamma)-r_k(\gamma\rightarrow \alpha)$) with respect to $x$ .}}
 \label{m5}
\end{center}
\end{figure}

An important point to focus on initially is the conceptual difference between this section and the previous one. In this section, we endeavor to interpret and study the dynamic sequence of a transfer process from a small black hole to a large black hole during a first-order phase transition, utilizing Kramer's escape rate. This rate represents the escape velocity of particles under thermal fluctuations from a known local minimum in the potential.\\
We will also examine whether conditions could arise that reverse the transfer process from a small to a large black hole. Typically, we expect a transition from an unstable local minimum to a stable global minimum. However, the question arises whether there is a range in which, due to environmental conditions such as temperature or pressure changes, the likelihood of a reverse process, namely a transfer from a large black hole to a small black hole, precedes the primary and natural process.\\For this particular scenario, we focus our attention on the frame-by-frame phase transition sequence based on the escape rate. An important point to note is that we are studying the dynamics, not the statics, of the frames. These frames are temporally continuous, meaning each frame at a given moment possesses its own physical concepts. However, these concepts must be interpretable in a way that they align with the overall understanding of the preceding and succeeding frames, ultimately corroborating the general perspective.\\In Fig.(\ref{m5}), we approach a state at the onset of phase transition where the chaotic process of phase transition has formed due to primary reasons, meaning the $\gamma$  minimum has found its initial form (as shown in Fig.(\ref{5b}), although, it has not yet become a global minimum. The small black hole (in the local minimum potential $\alpha$) has begun its intense movement towards the large black hole, considering the formation of a secondary minimum and prevailing conditions, which is clearly observed in Fig.(\ref{5c}). The probability of transition $\alpha\rightarrow\gamma$ is significantly higher than the other transition, Fig.(\ref{5c}), which is why the dominant transfer seems to be entirely from $\alpha\rightarrow\gamma$.  Since we are at the beginning of this transition, it seems that the big black hole has not yet formed, or in other words, it lacks enough particles for the formation of reverse processes. Therefore, as it is evident from the Fig.(\ref{5d}), the structure is under the complete control of the transition from a small black hole to a large one, and $\Delta r_k$ is completely positive.

\begin{figure}[H]
 \begin{center}
 \subfigure[]{
 \includegraphics[height=4.5cm,width=4cm]{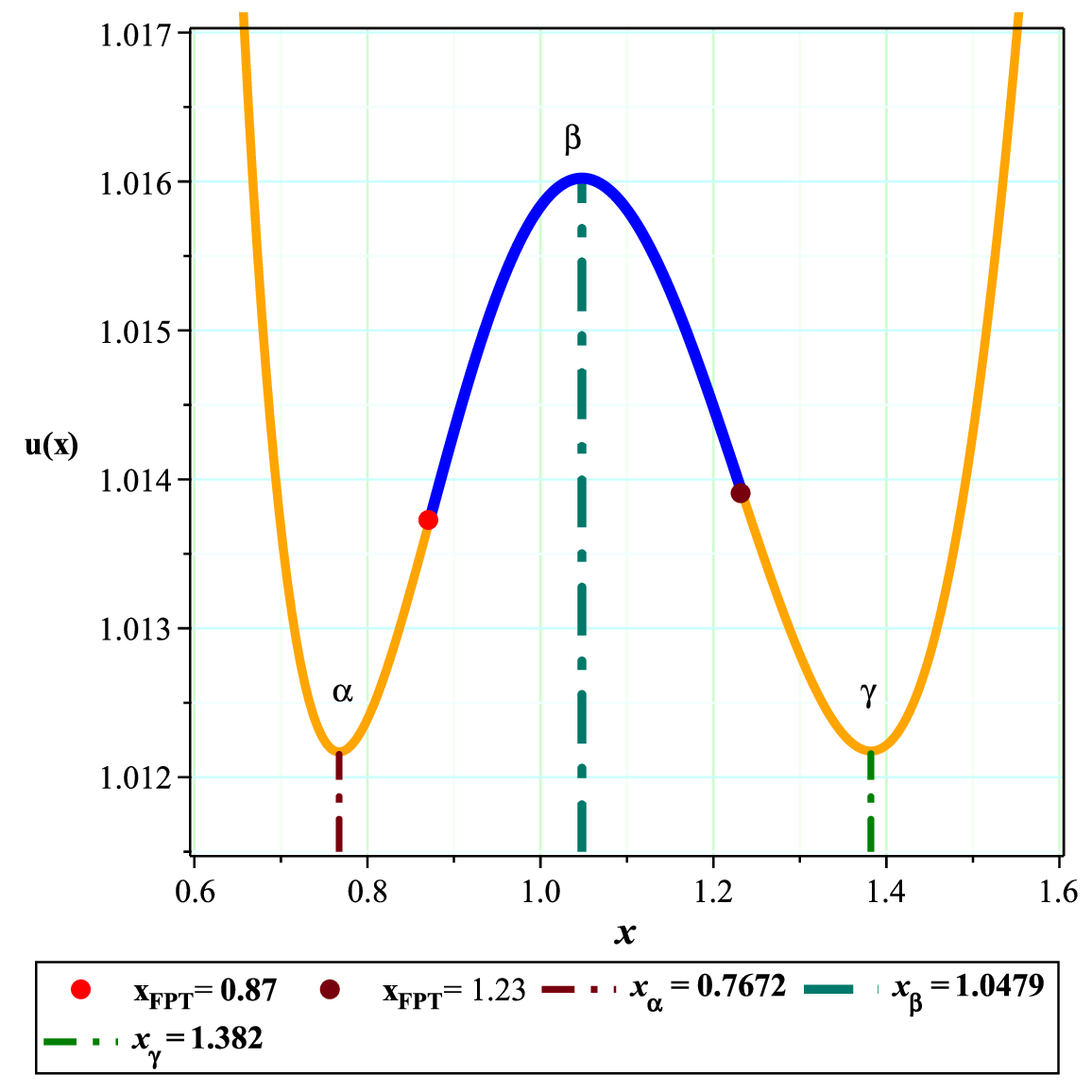}
 \label{6a}}
 \subfigure[]{
 \includegraphics[height=4.5cm,width=4cm]{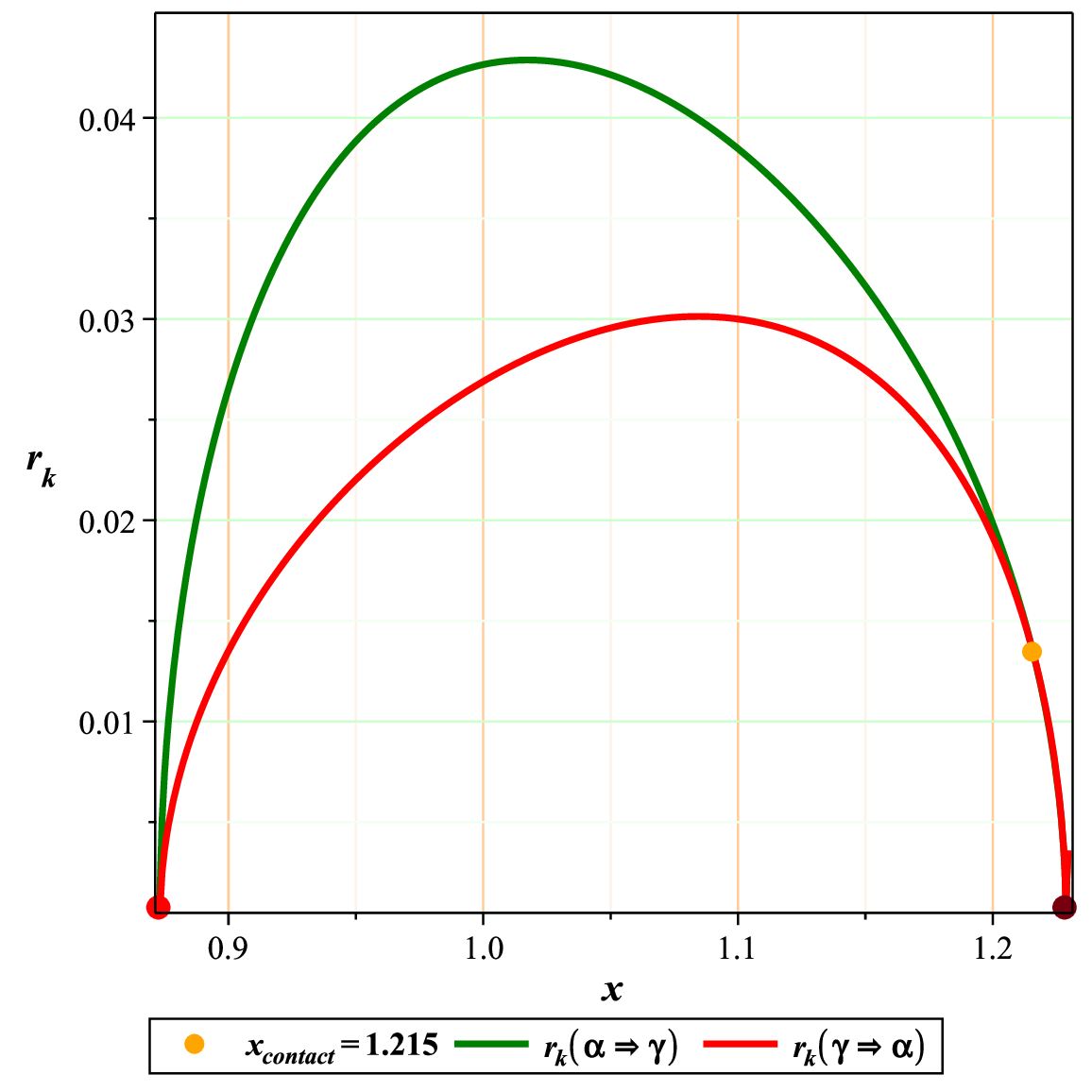}
 \label{6b}}
 \subfigure[]{
 \includegraphics[height=4.5cm,width=4cm]{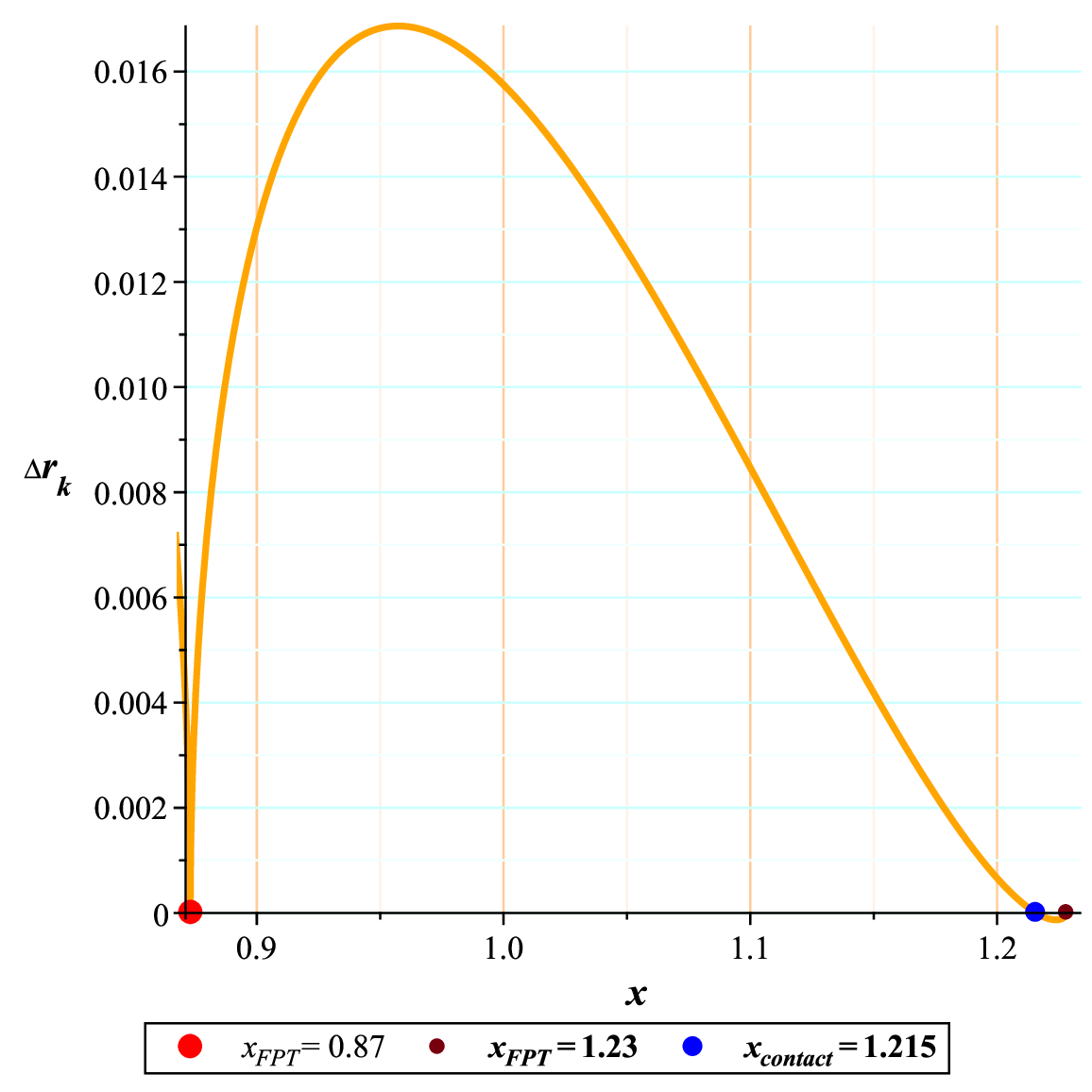}
 \label{6c}}
 \subfigure[]{
 \includegraphics[height=4.5cm,width=4cm]{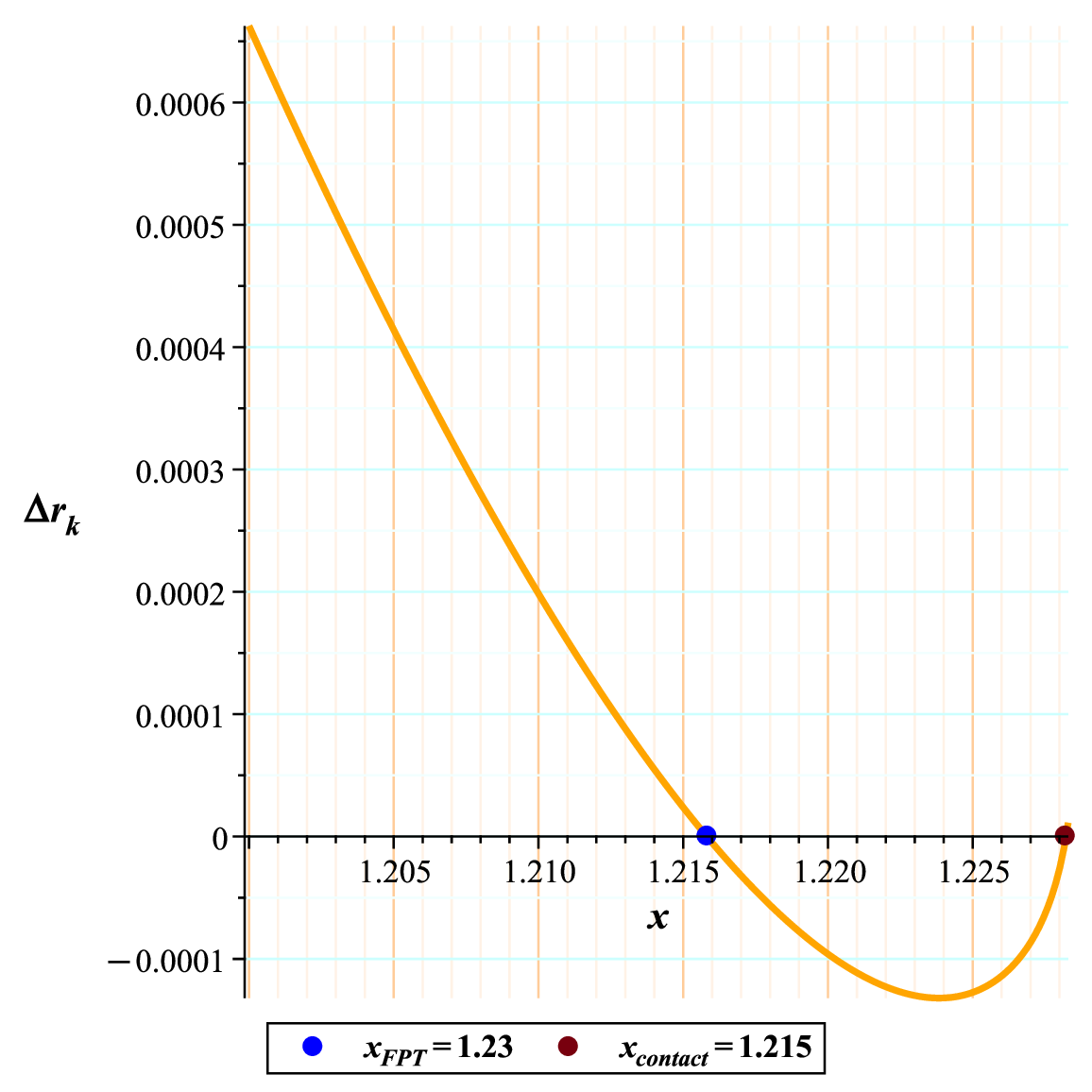}
 \label{6d}}
 \caption{\small{(\ref{6a}): The $u(r)$ against x for free parameters and the FPT points, \hspace{0.1cm} (\ref{6b}): comparing $r_k$ with respect to $x$ for($\alpha\rightarrow\gamma$) and ($\gamma\rightarrow\alpha$) and the coordinate of contact point \hspace{0.1cm} (\ref{6c}):$\Delta r_k$  with respect to $x$ \hspace{0.1cm} (\ref{6d}): The enlarged detail of the right end of the Fig (\ref{6c})}}
 \label{m6}
\end{center}
\end{figure}
In the subsequent stage, we examine a state midway through the phase transition, where the minimum $\gamma$  has assumed a relatively complete form (as depicted in Fig.(\ref{6a}), and although it has not yet become a global minimum, it is on par in depth with the alpha minimum.
If we were to consider this statically and in a single-frame format, it would appear that the escape rate and the probability of movement from the local beta maximum in both directions (i.e., towards the small or the large black hole) should be equal. However, it is important to note that on one hand, the environmental conditions imposing the phase transition on the system, and on the other hand, the shrinking radius —which is somewhat contrary to normal physical laws— still, make the probability of transition $\alpha\rightarrow \gamma$ greater than the other transition, as shown in Fig.(\ref{6b}). For this reason, the dominant transfer seems to continue to be from $\alpha\rightarrow \gamma$.
Nevertheless, as is evident in Fig. (\ref{6b}), a collision point has formed between the escape rate diagrams, and beyond this contact point, it is clear that the reverse process, albeit in a very small region at the end of the diagram, predominates over the direct process. It appears that as we approach the end of the phase transition, a negative region (Figs. (\ref{6c}) and (\ref{6d}) is forming which somehow, by reinforcing feedback processes, sets the stage for a change in the initial conditions and ultimately halts the phase transition.

\begin{figure}[H]
 \begin{center}
 \subfigure[]{
 \includegraphics[height=4.5cm,width=5cm]{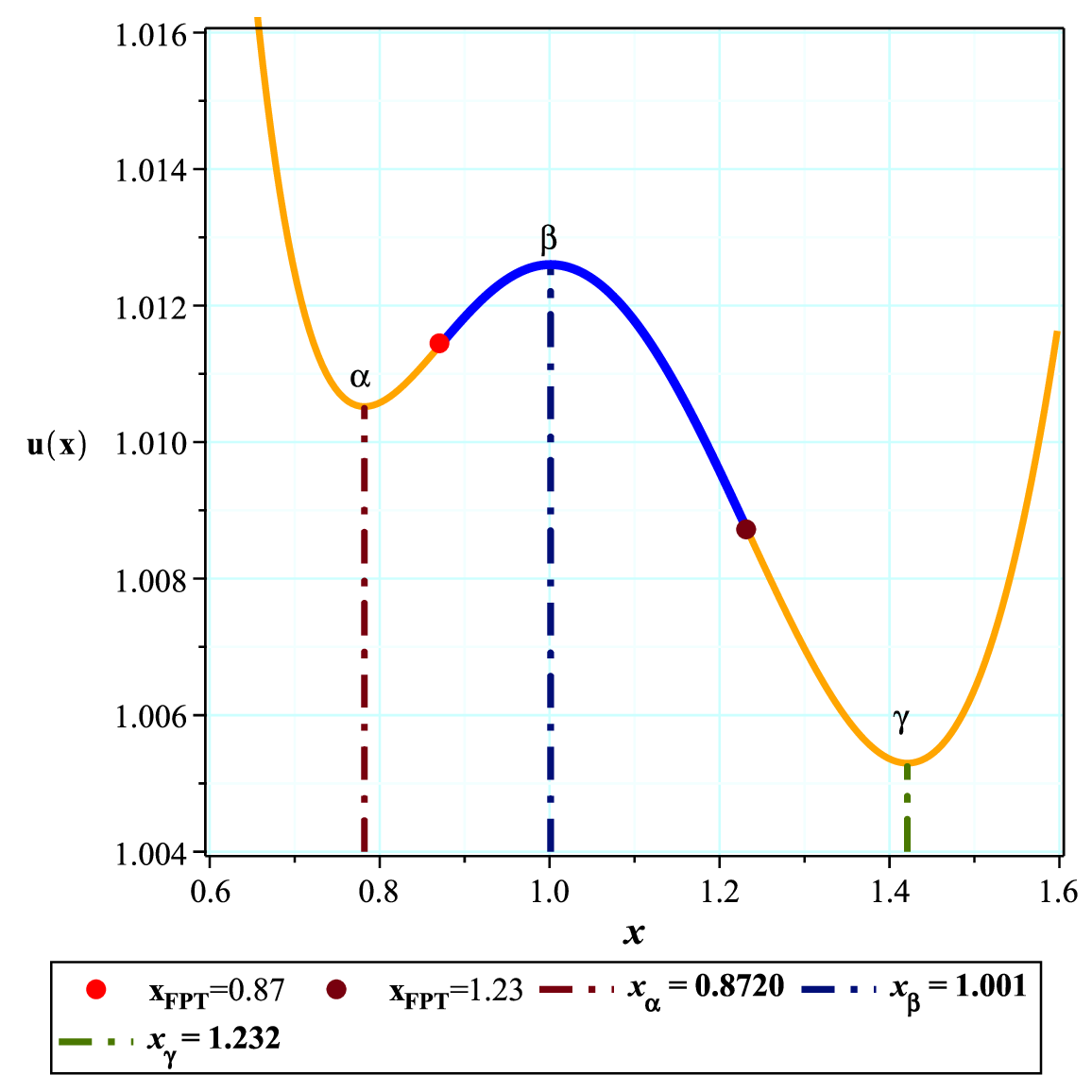}
 \label{7a}}
 \subfigure[]{
 \includegraphics[height=4.5cm,width=5cm]{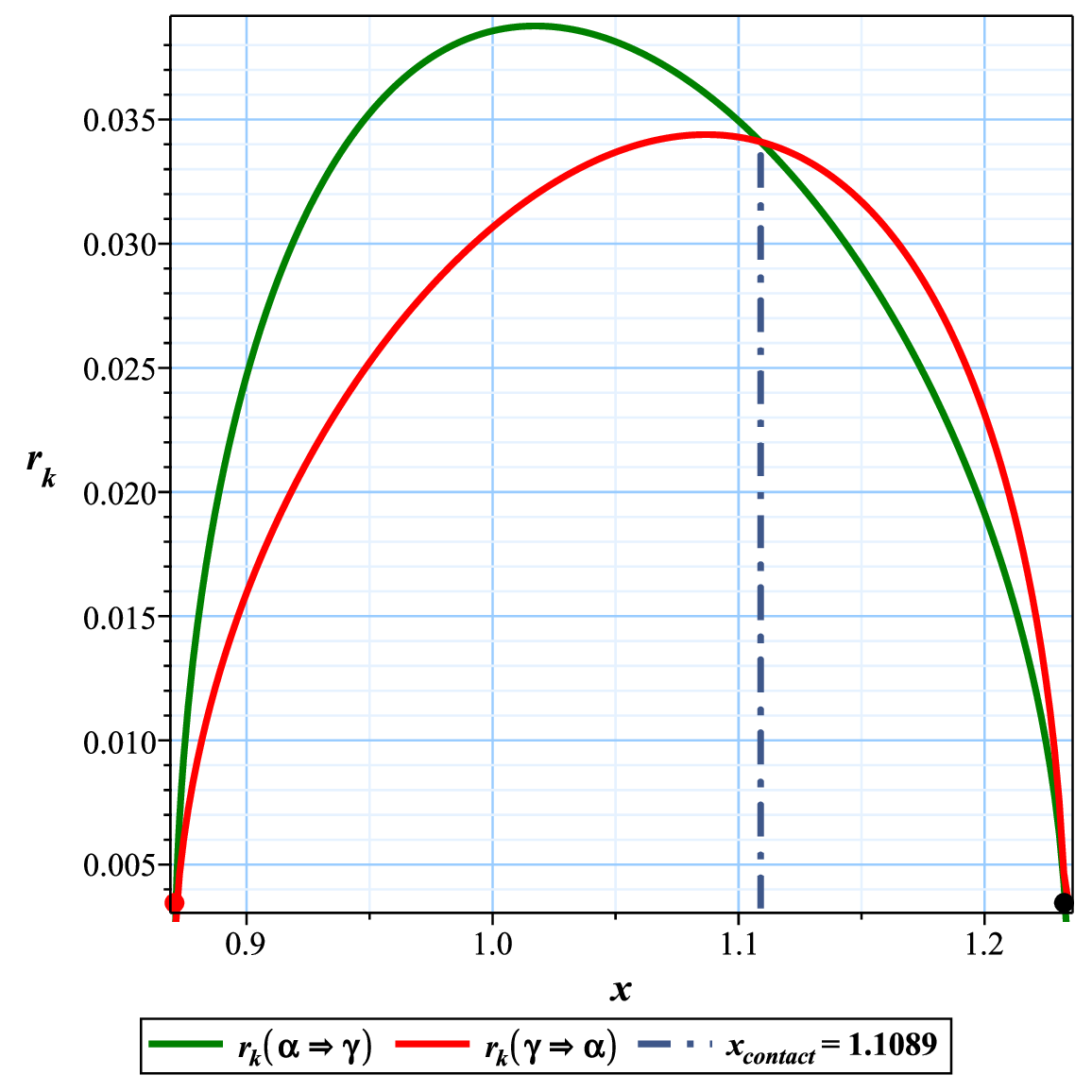}
 \label{7b}}
 \subfigure[]{
 \includegraphics[height=4.5cm,width=5cm]{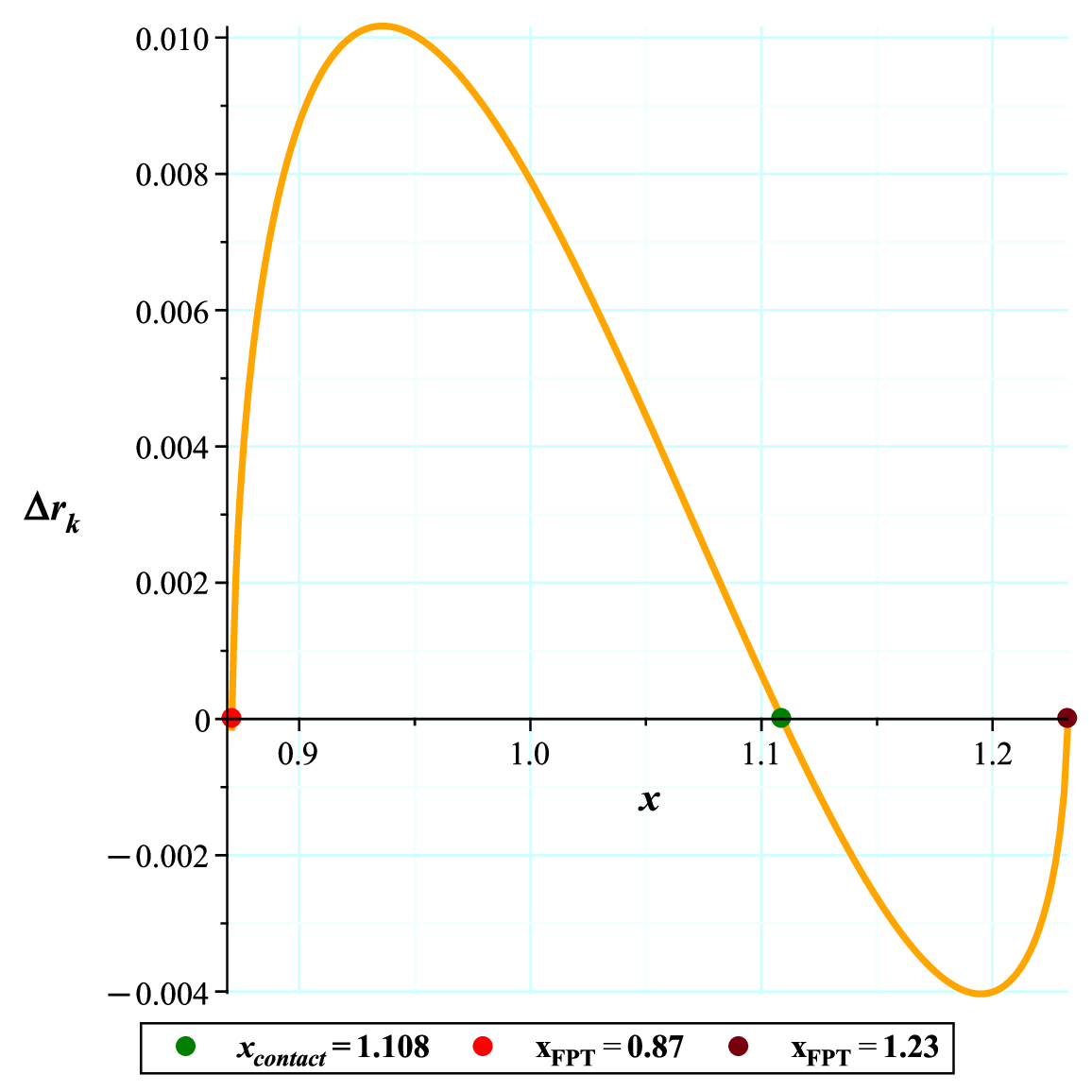}
 \label{7c}}
 \caption{\small{(\ref{7a}): The $u(r)$ against x for free parameters and the FPT points, \hspace{0.1cm} (\ref{7b}): comparing $r_k$ with respect to $x$ for($\alpha\rightarrow\gamma$) and ($\gamma\rightarrow\alpha$) and the coordinate of contact point \hspace{0.1cm} (\ref{7c}):$\Delta r_k$  with respect to $x$ .  }}
 \label{m7}
\end{center}
\end{figure}

Ultimately, we turn our attention to a state near the end of the phase transition pathway, where the $\gamma$ has taken on a relatively complete shape (as illustrated in Fig.(\ref{7a}) and has become a global minimum.
Although the environmental conditions that are imposing the phase transition on the system continue to make the probability of transition $\alpha \rightarrow \beta$ greater than the other transitions, as depicted in Fig.(\ref{7b}). However, as can be seen from Fig. (\ref{7b}), the coordinates of the contact point have moved significantly upwards, to the extent that the probability of the reverse process cannot be ignored in comparison to the direct process, and now a larger region is dominated by this process, as shown in Fig.(\ref{7c}). As we observed throughout this sequence, the closer we got to the end of the phase transition, the contact point advanced towards, increasing the probability of the reverse transition. It seems that this reverse process may, in some way, be the system's reaction to prevent the phase transition process from becoming uncontrollable and the radial growth that might overshadow the stability of the black hole. It may also be hypothesized that the black hole, to prevent the phase transition process from becoming uncontrollable, instead of a direct and sudden transition from the small black hole to the large black hole (akin to a completely damped oscillation), performs this process in the form of a slowly damped, quasi-oscillatory manner.
\section{4D AdS Einstein-gauss-bonnet-Yang-Mills black hole with a cloud of strings}
Since we tried to explain the contents as much as possible in the previous sections, considering the computational similarities, in this section, we will only mention the basic relationships, and we will bring the similar and unnecessary diagrams in Appendix A. The second selected model is the (EGB-YM-CS) black hole. So the metric of 4D AdS Einstein-gauss-bonnet-yang-mills black hole with a cloud of strings is as follows\cite{52},
\begin{equation*}\label{27}
f \! \left(r \right)=1+\frac{r^{2} \left(1-\sqrt{1-4 g \left(\frac{8 \pi  P}{3}+\frac{q^{2}}{r^{4}}-\frac{2 M}{r^{3}}+\frac{a}{r^{2}}\right)}\right)}{2 g}.
\end{equation*}

For the main quantities of this model i.e. mass $M$, Hawking temperature $T_H$, entropy $S$, pressure $P$ and volume $V$ we will have\cite{52},
\begin{equation}\label{28}
M =\frac{8 r^{4} P \pi +\left(3 a +3\right) r^{2}+3 q^{2}+3 g}{6 r},
\end{equation}
\begin{equation}\label{29}
T_H =\frac{8 r^{4} P \pi +r^{2} \left(a +1\right)-q^{2}-g}{4 \left(r^{2}+2 g \right) r \pi},
\end{equation}
\begin{equation}\label{30}
S =r^{2} \pi +4 \ln \! \left(r \right) g \pi ,
\end{equation}
\begin{equation}\label{31}
P =\frac{4 \pi  T \,r^{3}+8 \pi  T g r -r^{2} a +q^{2}-r^{2}+g}{8 r^{4} \pi},
\end{equation}
\begin{equation}\label{32}
V =\frac{4 \pi  r^{3}}{3},
\end{equation}
where q is the charge of the black hole, "a" is related to the density of clouds of strings and comes from the energy-momentum tensor of the clouds of strings and $g$ \footnote{ Gauss-Bonnet coupling constant is displayed with g instead of $\alpha$ to avoid mistakes with clouds of strings parameter} is a positive Gauss-Bonnet coupling constant.
\subsubsection{ The landscape free energy}
According to Eqs. (\ref{2}), (\ref{3}), (\ref{4}) and (\ref{7}) for Gibbs and Landau free energy function and  thermal potential of this model, we will have,
\begin{equation}\label{33}
\textbf{G} =\frac{4 P \pi  r^{3}}{3}+\frac{a r}{2}+\frac{g}{2 r}+\frac{r}{2}+\frac{q^{2}}{2 r}+\frac{\left(-8 r^{4} P \pi -r^{2} \left(a +1\right)+q^{2}+g \right) \left(4 \left(\frac{2 \left(r -1\right)}{r +1}+\frac{2 \left(r -1\right)^{3}}{3 \left(r +1\right)^{3}}\right) g +r^{2}\right)}{4 \left(r^{2}+2 g \right) r},
\end{equation}
\begin{equation}\label{34}
G_{L}=\frac{4 P \pi  r^{3}}{3}+\frac{a r}{2}+\frac{g}{2 r}+\frac{r}{2}+\frac{q^{2}}{2 r}-\left(4 g \ln \! \left(r \right)+r^{2}\right) T \pi,
\end{equation}
\begin{equation}\label{35}
\textbf{U} =\frac{8 r^{4} P \pi +\left(3 a +3\right) r^{2}+3 q^{2}+3 g}{6 r}-\left(r^{2} \pi +4 \ln \! \left(r \right) g \pi \right) T ,
\end{equation}
\begin{equation}\label{36}
\textbf{L} =P X -\frac{2 \,6^{\frac{2}{3}} \left(-\frac{3 \pi^{\frac{1}{3}} 6^{\frac{2}{3}} \left(a +1\right) X^{\frac{2}{3}}}{16}+\pi^{\frac{5}{3}} T g 6^{\frac{1}{3}} \ln \! \left(X \right) X^{\frac{1}{3}}+\frac{9 \pi  \left(T X -\frac{2 q^{2}}{3}-\frac{2 g}{3}\right)}{8}\right)}{9 \pi^{\frac{2}{3}} X^{\frac{1}{3}}} ,
\end{equation}
Now we will go to a special state that is formed based on the arbitrary parameter setting. For this purpose, by considering the values  $ a=1.5, q=0.5, g=0.5 $ for the parameters and with a little calculation for the critical quantities, we will have,
\begin{equation}\label{37}
\begin{split}
&T_{c}=  0.0978,\hspace{0.7cm} P_{c} =  0.0075,\hspace{0.7cm} r_{c} =  2.269260985,\\
&V_{c}= 48.94878867,\hspace{0.7cm} U_{c} = 1.28320692868,\hspace{0.7cm} G_{c} =  1.28662690146,\\
\end{split}
\end{equation}
In this model, we still use Eq.(\ref{17}) to make the equations dimensionless.
\subsubsection{ Kramer’s escape rate }
In order to examine the escape rate in this black hole, given that the motion trajectory is analogous to the previous state, we will delineate only two scenarios to prevent redundancy: one at the onset of phase transition and the other in proximity to the conclusion of phase transition. We endeavor to eschew similar descriptions as much as possible.
\begin{figure}[H]
 \begin{center}
 \subfigure[]{
 \includegraphics[height=4.5cm,width=4cm]{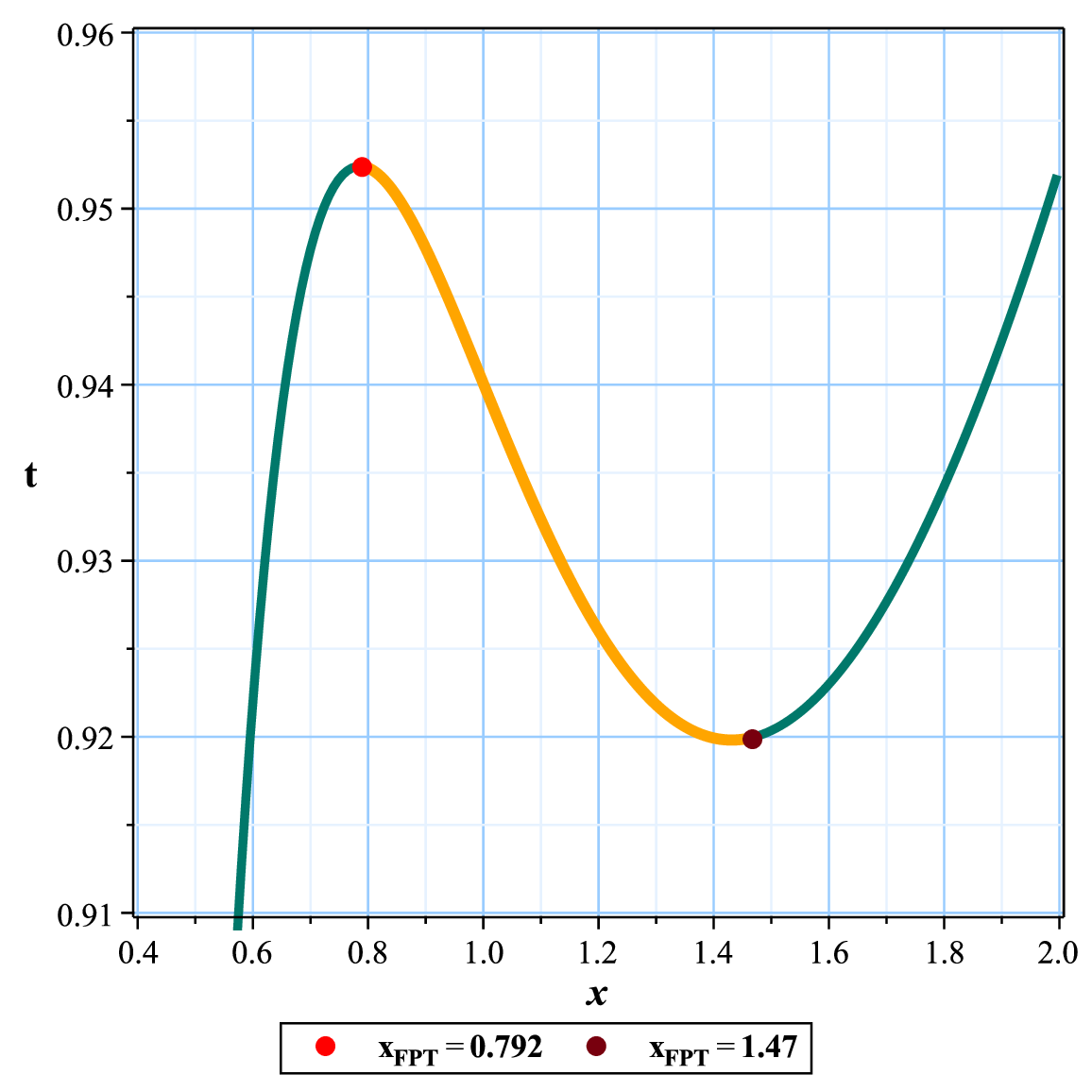}
 \label{8a}}
 \subfigure[]{
 \includegraphics[height=4.5cm,width=4cm]{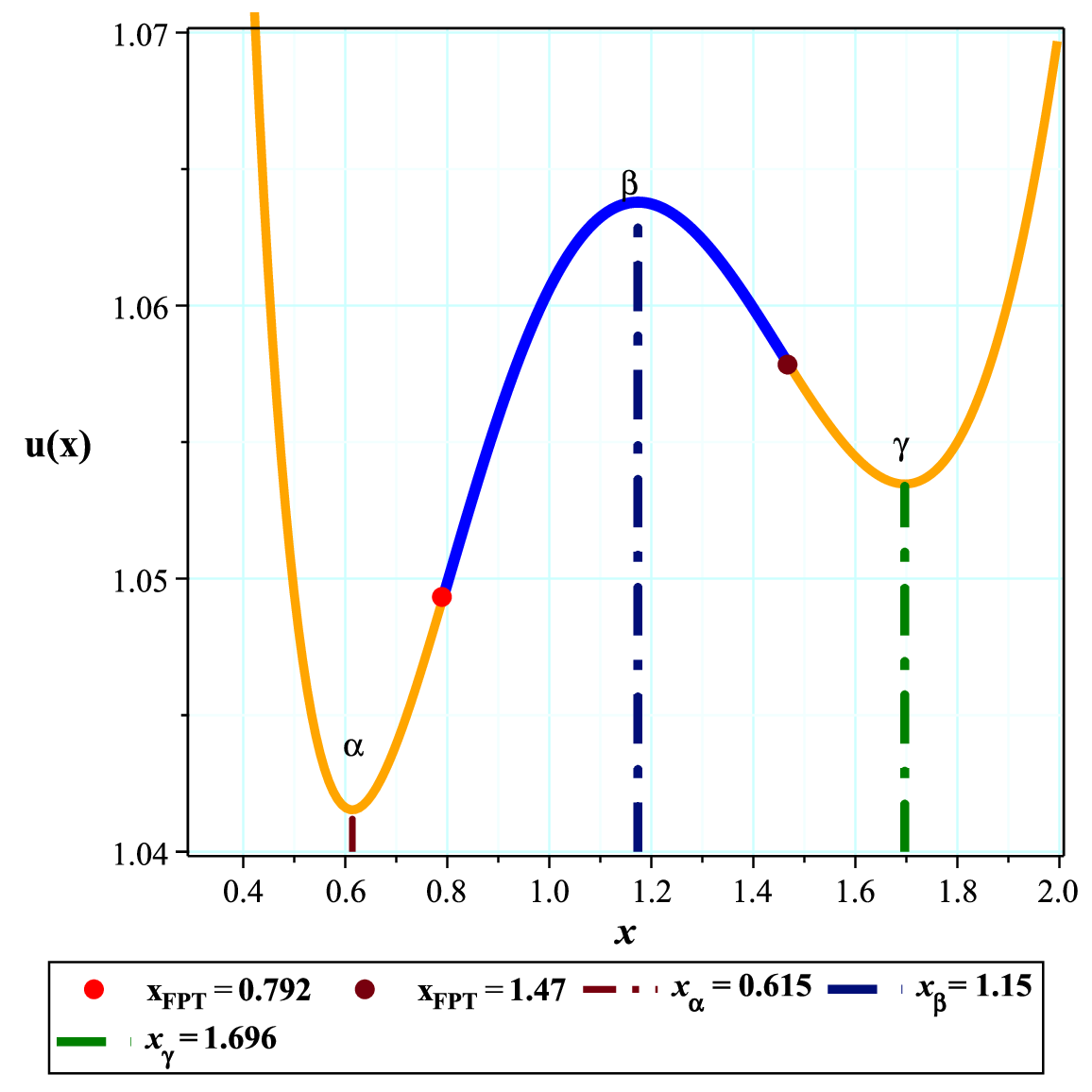}
 \label{8b}}
 \subfigure[]{
 \includegraphics[height=4.5cm,width=4cm]{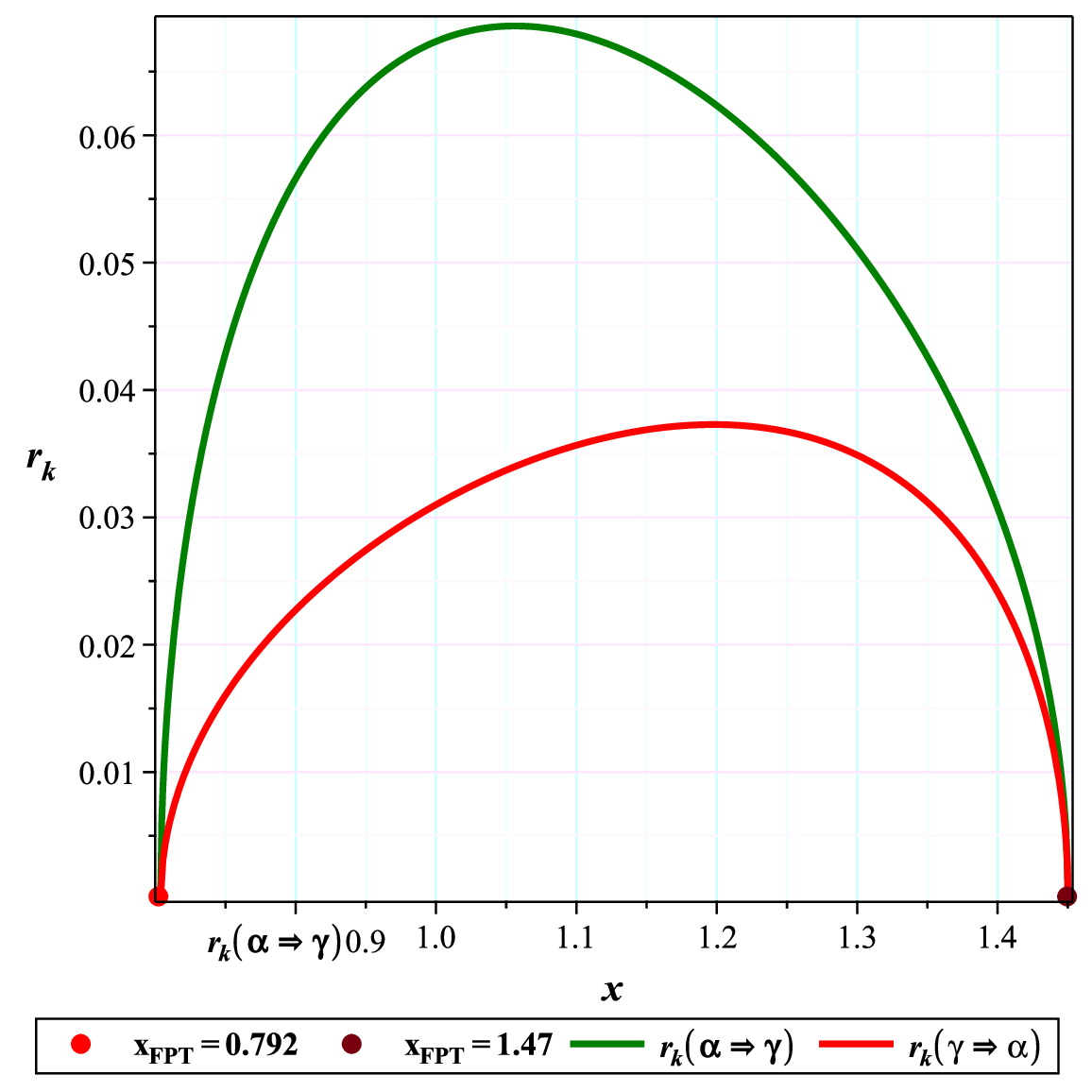}
 \label{8c}}
 \subfigure[]{
 \includegraphics[height=4.5cm,width=4cm]{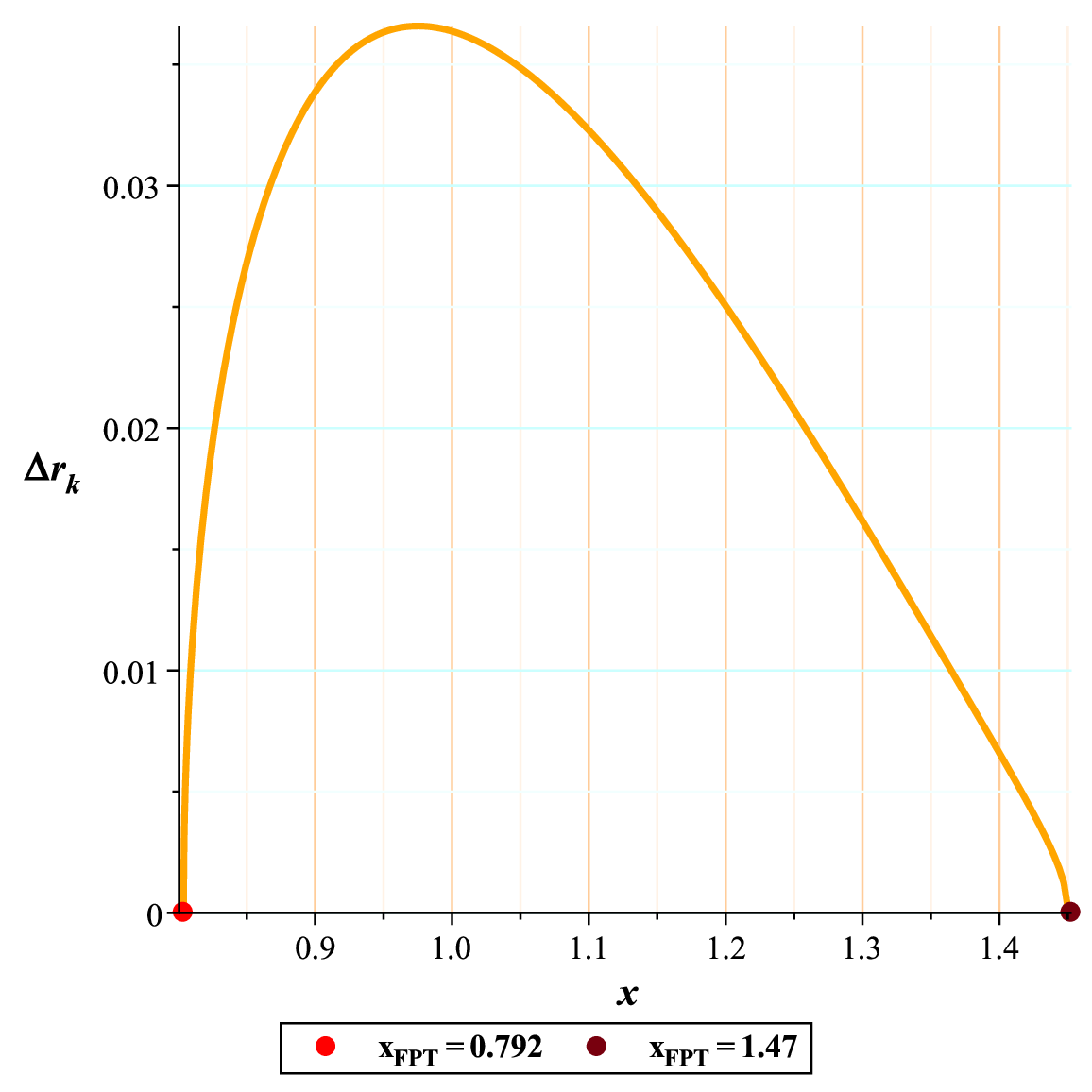}
 \label{8d}}
   \caption{\small{t versus x  and the First order Phase Transition region (FPT) in Fig. (\ref{8a}),\hspace{0.1cm} (\ref{8b}): The graph of $u(r)$ against x for free parameters and the FPT points, \hspace{0.1cm} (\ref{8c}): comparing $r_k$ with respect to $x$ for($\alpha\rightarrow\gamma$) and ($\gamma\rightarrow\alpha$) and the coordinate of contact point \hspace{0.1cm} Fig. (\ref{8d}): $\Delta r_k$  with respect to $x$ .}}
 \label{m8}
\end{center}
\end{figure}
As evident from Fig.(\ref{8d}), just like before, at the initial of phase transition of the process is predominantly characterized by a  direct transition. As observed in the diagram, the positive region is effectively the dominant area and no reactive structure appears until the end of the process.
\begin{figure}[H]
 \begin{center}
 \subfigure[]{
 \includegraphics[height=4.5cm,width=5cm]{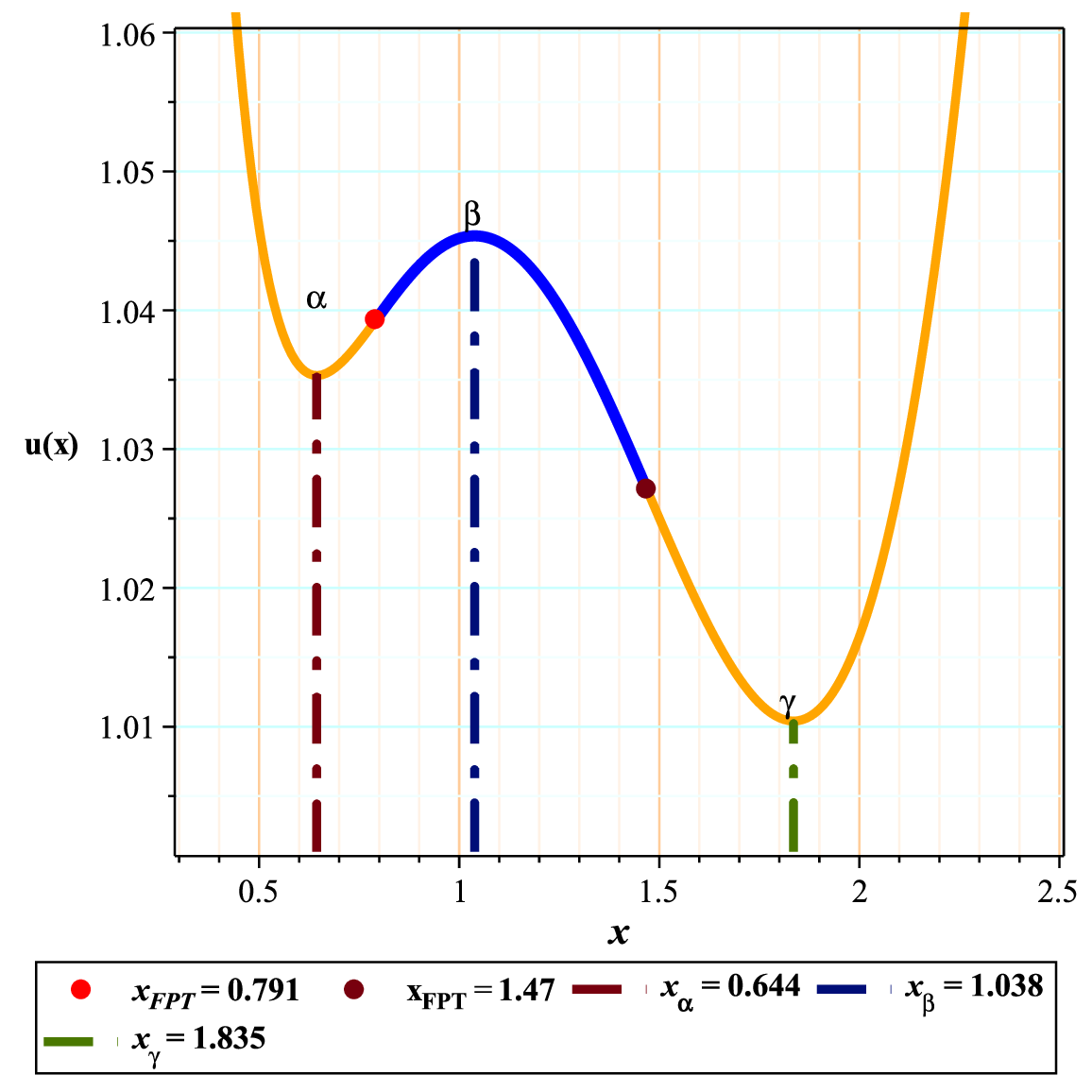}
 \label{9a}}
 \subfigure[]{
 \includegraphics[height=4.5cm,width=5cm]{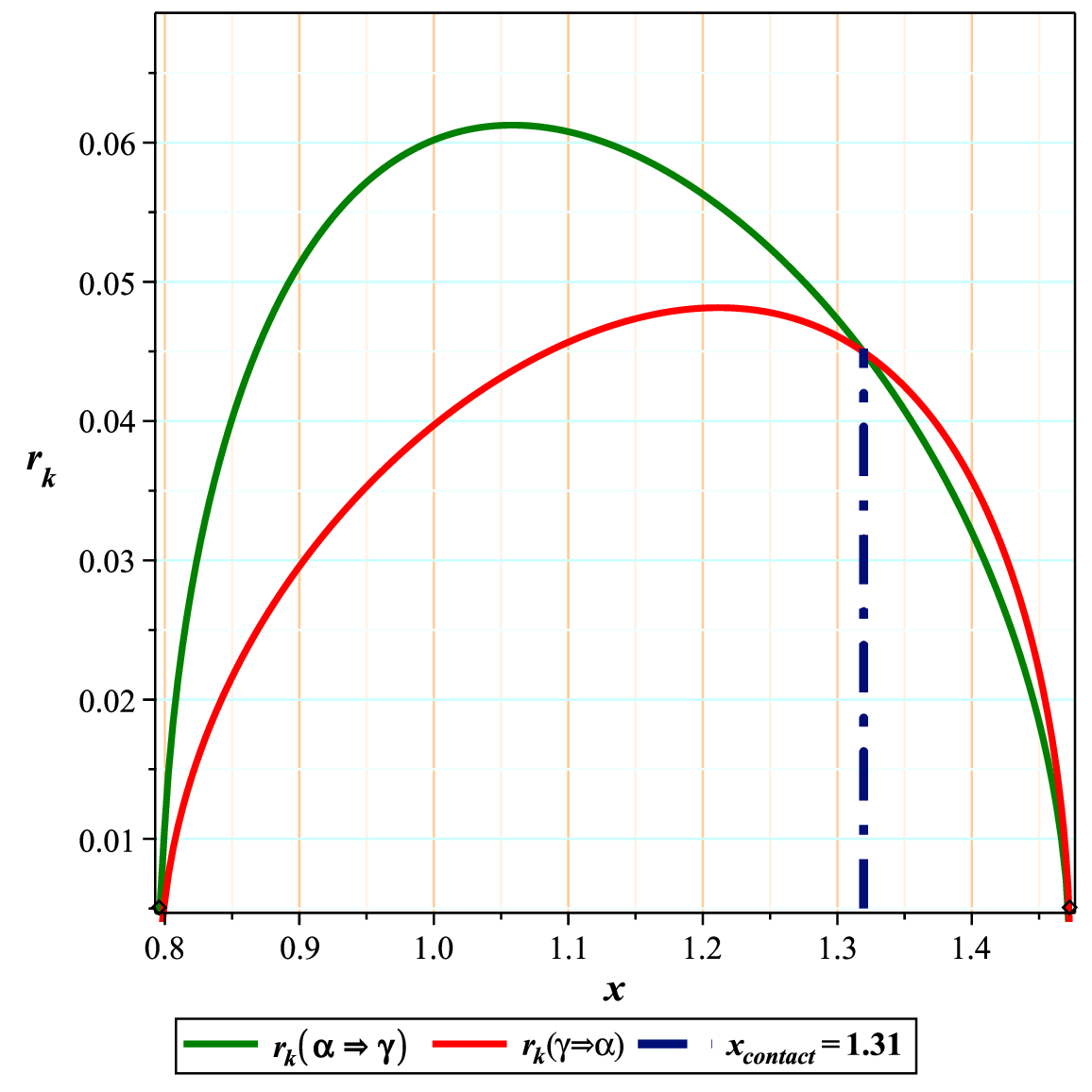}
 \label{9b}}
 \subfigure[]{
 \includegraphics[height=4.5cm,width=5cm]{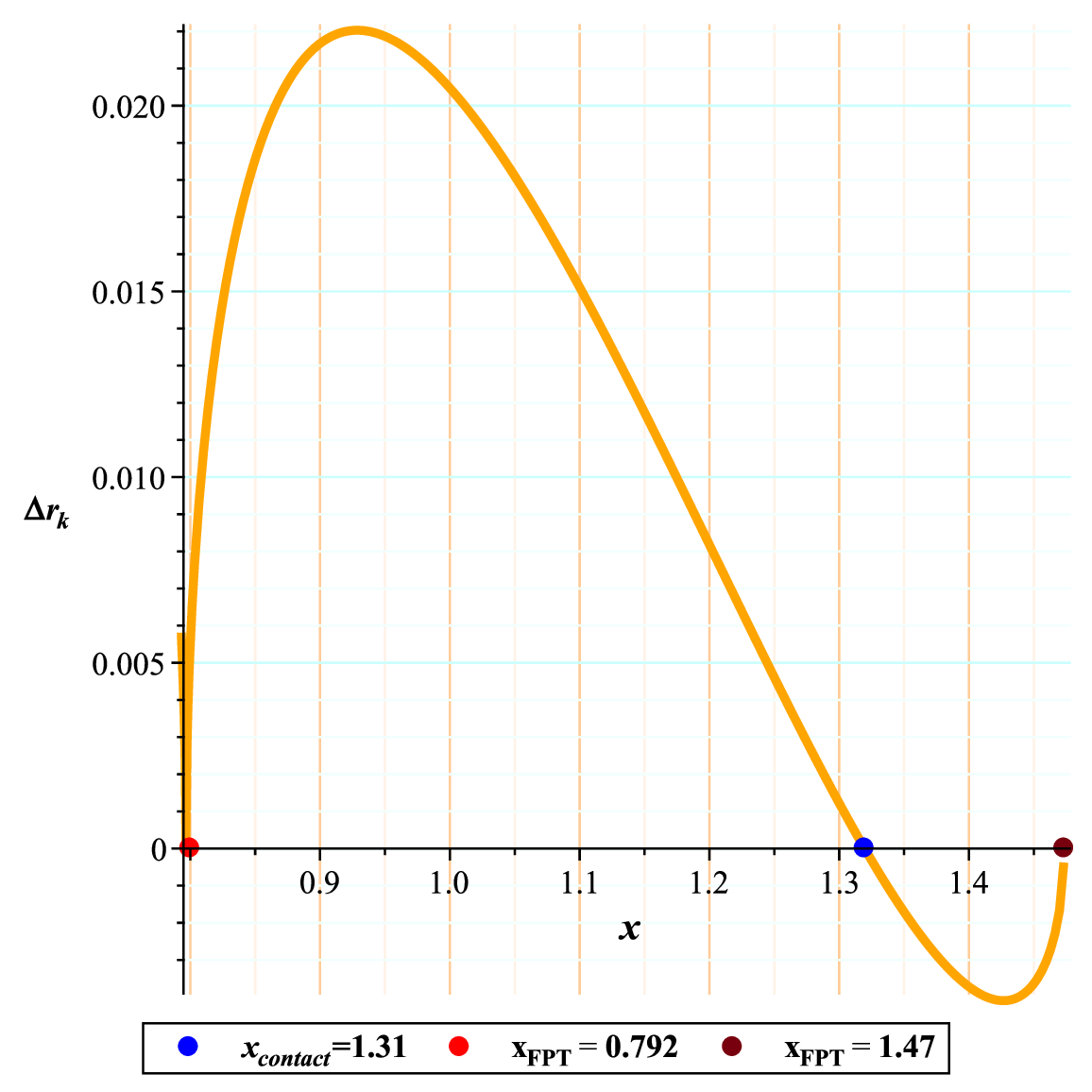}
 \label{9c}}
   \caption{\small{(\ref{9a}): The $u(r)$ against x for free parameters and the FPT points, \hspace{0.1cm} (\ref{9b}): comparing $r_k$ with respect to $x$ for($\alpha\rightarrow\gamma$) and ($\gamma\rightarrow\alpha$) and the coordinate of contact point \hspace{0.1cm} (9c):$\Delta r_k$  with respect to $x$ . }}
 \label{m9}
\end{center}
\end{figure}
As illustrated in Fig.(\ref{m9}), with the progression of the phase transition process and as we approach the end, although the primary contribution is still from a direct transition, as depicted in Fig.(\ref{9d}), but the negative region has undergone significant growth and has become comparable to the positive region.
 It appears that, similar to the previous case, the reactive processes have become more potent in near the end of the process.
\section{Discussion and comparison}
In this section, for a better understanding, comparison, and interpretation of the results, it is first beneficial to delve a little into the concept of transition from small to large black holes (or large to small). The transition from a small to a large black hole can generally occur through the accretion of matter and energy from the surrounding environment of the black hole, or it can also occur through the merging of black holes, which may be accompanied by intensified emission of gravitational waves.
However, from a thermodynamic perspective, this transition seems to encompass any physical process that facilitates changes in temperature, pressure, and other thermodynamic parameters in such a way that it leads to an increase in the radius of the event horizon and consequently, the enlargement of the black hole, which, given the super-gravitational structure of black holes, is a highly probable process.
The reverse process, the transition from a large to a small black hole, can occur through processes such as Hawking radiation, where black holes emit particles due to quantum effects near the event horizon. Over time, this radiation can lead to a reduction in the mass and size of the black hole and its eventual evaporation. From a thermodynamic perspective, this phenomenon is equivalent to any process that creates a change in thermodynamic parameters leading to a reduction in the radius of the horizon and the shrinking of the black hole. As observed in Figs. (\ref{5b}),(\ref{6a}),(\ref{8b}), (\ref{7a}) and (\ref{9a}), our black hole initially resides in a stable local minimum,$\alpha$, which globally is considered an unstable minimum. Consequently, as one would naturally expect, with the slightest thermal perturbation, and corresponding pressure, the system increasingly tends to transition towards the global minimum,$ \gamma$, signifying an increase in radius. In both the NLM-C-Q-PFD and EGB-YM-CS models, upon examining the escape rate diagrams, we observe that once the radius exceeds a certain threshold, a small domain emerges where the reverse process forms and prevails. In other words, with the dominance of $r_k(\gamma\rightarrow\alpha)$ over $r_k(\alpha\rightarrow\gamma)$, a condition arises allowing black holes to transition from $\gamma$ to $\alpha$. Notably, over time and the sequence of frames, this seemingly negligible and very small region at the onset of phase transition gradually transforms into a significant area in the final frames.
At first glance, the transition from $\gamma$ to $\alpha$ may seem less compatible with physical conditions because, within the same energy structure and without topological changes in the energy function (i.e., maintaining the gamma state as the global minimum and alpha as the local minimum), the system prefers to revert to the ( $\alpha$ ) state with a smaller radius, leading to a decrease in entropy and an increase in energy level. This situation can be termed 'improbable' or 'less likely' from a physical standpoint.
However, on the other hand, from a Kramers' perspective on the reverse process, it can be stated that particles situated in a potential well under thermal fluctuations can always escape with a specific probability coefficient (which is a function of the potential depth). Therefore, from the perspective of particles in the gamma potential, this movement is not considered an improbable phenomenon. Moreover, if we consider the transition from a smaller to a larger black hole as a general action, a natural reaction should form against it. The increase in radius and the consequent decrease in energy level is a recognized form of this reaction, leading to the reality that the escape rate, after reaching a peak, undergoes a decreasing trend until the probability becomes zero. But does this zero probability occur at any desired radius?\\
These ideas, coupled with the observation that as we approach the terminal point of phase transition, we witness an increased probability of the reverse process, have led us to propose that perhaps this reverse trend is, in fact, a natural response of the black hole to phase transition. It may indeed be an effort to limit and control the phase transition to prevent excessive radial growth and uncontrolled instability, ensuring that the process of moving towards a larger black hole concludes under conditions that maintain the stability of the final structure and do not disrupt the viability of the black hole.
\section{Conclusion}
The study of black hole phase transitions based on the Gibbs free energy function has played a significant and influential role in advancing our understanding of black hole thermodynamics. However, a common issue with this conventional approach is that it seems to consider phase transitions only in a static form, leaving questions about the dynamics of phase transitions largely unanswered. The proposition that the dynamics of phase transitions in black holes are governed by thermodynamic and statistical forces has led to the possibility of examining phase transitions based on the stochastic Langevin equation or the Fokker-Planck equation \cite{21}, considering the 'mean first passage time'\cite{22,23,24,25,26,27,28,29,30,31,32,33,34,35,36,37,38,39,40,41,42,43,44}. In this context, the introduction of the Kramers escape rate in the study of first-order phase transitions\cite{48} prompted us to examine various free energy landscape forms and subsequently investigate the Kramers escape rate for the phase transitions of more complex black holes . These black holes are specifically formed based on cosmic and stringy features, meaning they interact more with the characteristics of dark structures. Our goal was to see what effect the addition of dark structure and stringy parameters would have on the dynamic behavior of phase transitions in this category of black holes. Accordingly, in this paper, we have dynamically investigated the transition from a small black hole to a large black hole in first-order phase transitions from the perspective of the Kramers escape rate. Considering the escape rate as a function of distance or black hole radius, we examined whether the dynamic behavior of this phase transition for the black hole under study is exactly as expected at all moments of the transition. That is, firstly, the escape rate initially increases with the radius from a local minimum towards a global minimum and, after reaching a certain maximum, gradually decreases due to the activation of a reaction mechanism until it tends to zero. Secondly, the rate of escape and leakage from a small black hole to a large black hole always follows a logical and direct trend. With this background, we studied the models (NLM-C-Q-PFD) and (EGB-YM-CS) and found that the escape rate diagram as a function of radius shows something different in specific regions. This means that the diagrams at the beginning, middle, and near the end of phase transition, at a specific radius, show equal probabilities of transition from a small to a large black hole and vice versa, which we have called contact coordinates in this paper. Beyond this radius, in a specific range up to the final point of phase transition, the reverse process becomes completely probable, meaning the probability of transitioning from a large black hole to a small black hole becomes dominant, and it seems that black holes prefer to travel from a large black hole to a small one. An interesting point about this region is that at the onset of phase transition, this area is very small, and as we approach the end in the sequence of frames, this area grows significantly. Since, on the one hand, the growth of the probability of the reverse process corresponds to the increase in radius and, on the other hand, it is in the direction of the end of the phase transition process, it seems that in fact, this reverse trend is a reactive mechanism by the black hole. It is actually an effort to limit and control the phase transition to prevent excessive radial growth and prevent any uncontrolled instability so that the black hole can conclude the process towards a larger black hole under conditions that preserve the stability of the final structure and do not disrupt the survival of the black hole.

\section{Appendix A: Behavioral sequence of the free energy landscape for (EGB-YM-CS) black hole}
\begin{figure}[H]
 \begin{center}
 \includegraphics[height=5cm,width=5cm]{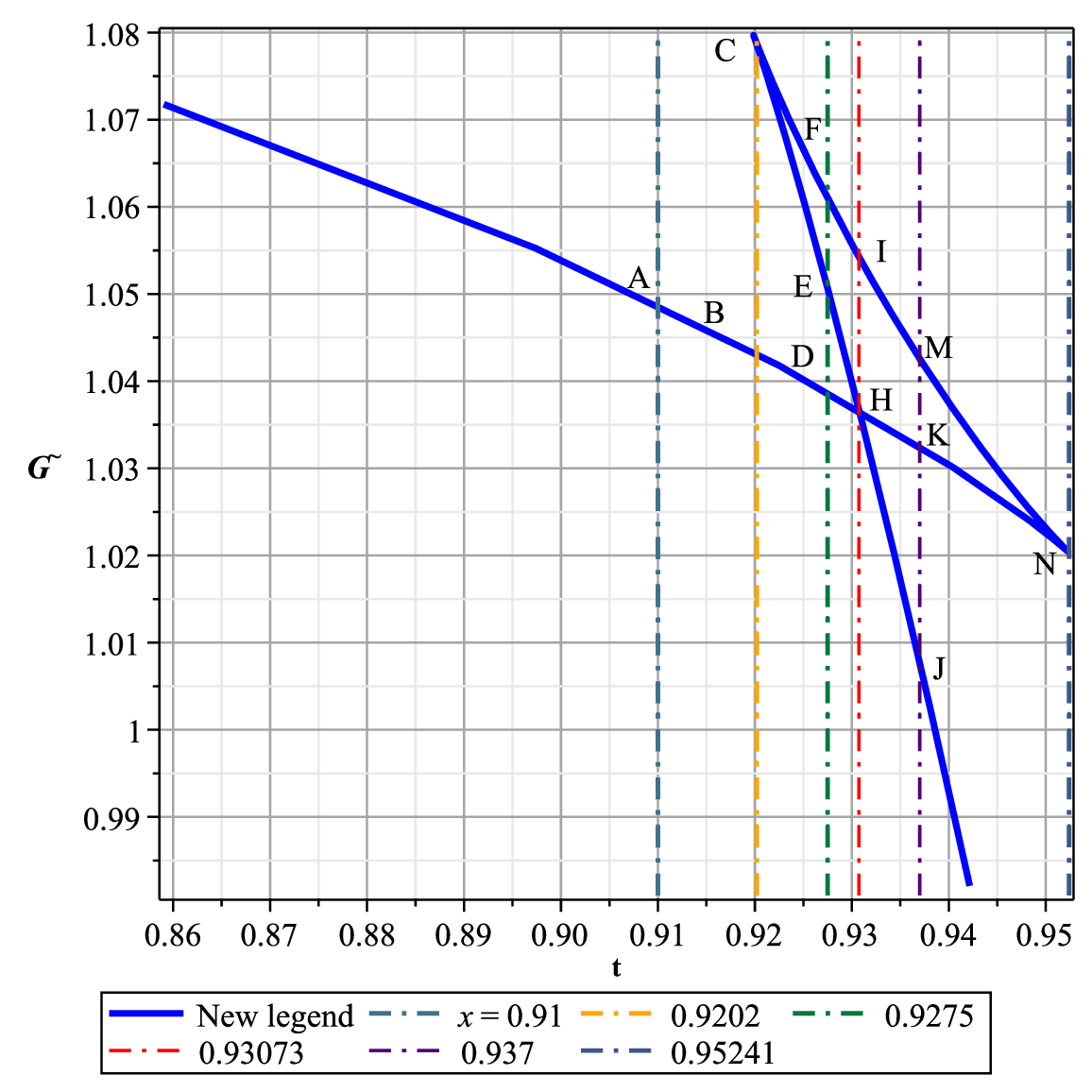}
 \caption{\small{The Gibbs free energy landscape in terms of t }}
 \label{m10}
\end{center}
\end{figure}

\begin{figure}[H]
 \begin{center}
 \subfigure[]{
 \includegraphics[height=4.15cm,width=4cm]{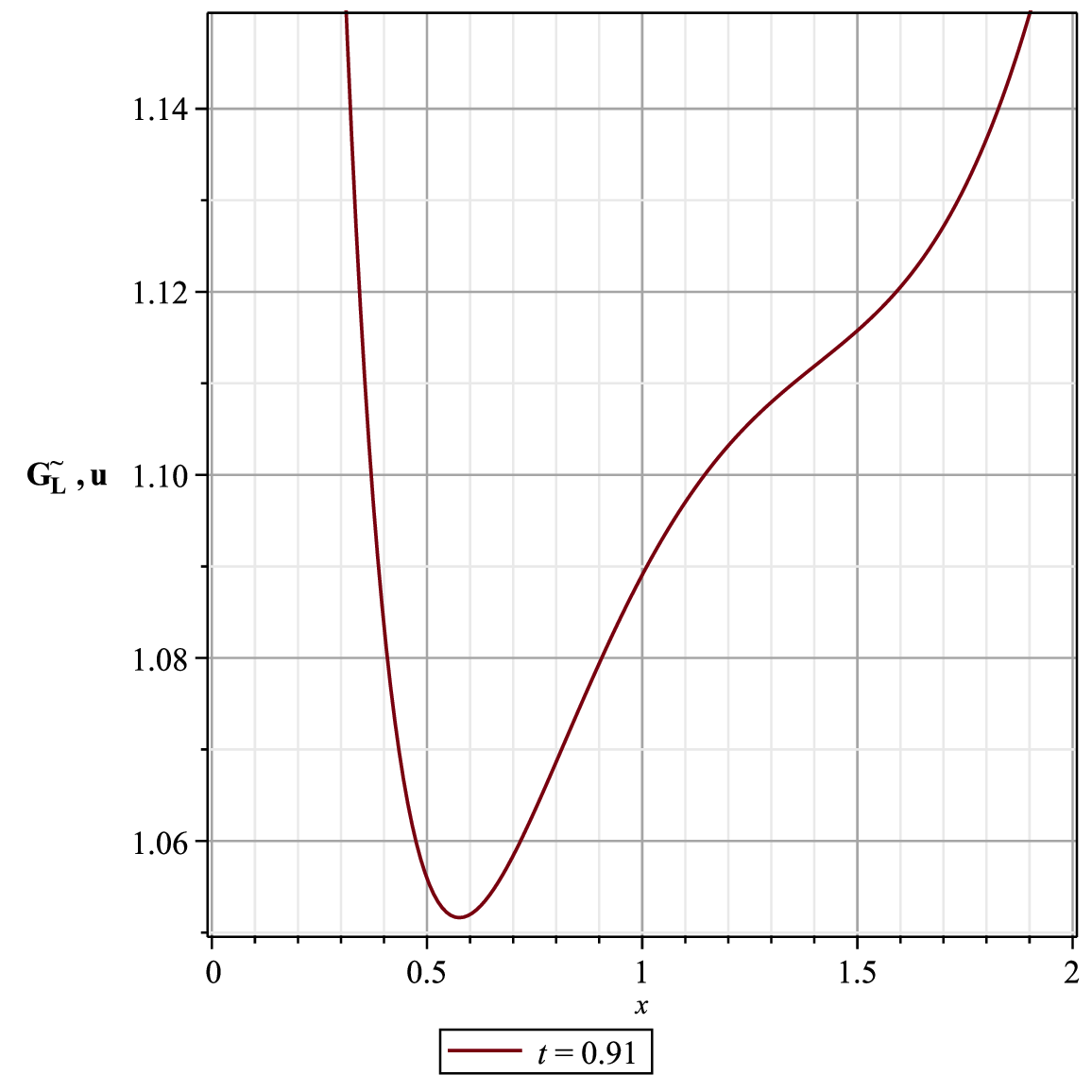}
 \label{11a}}
 \subfigure[]{
 \includegraphics[height=4.20cm,width=4cm]{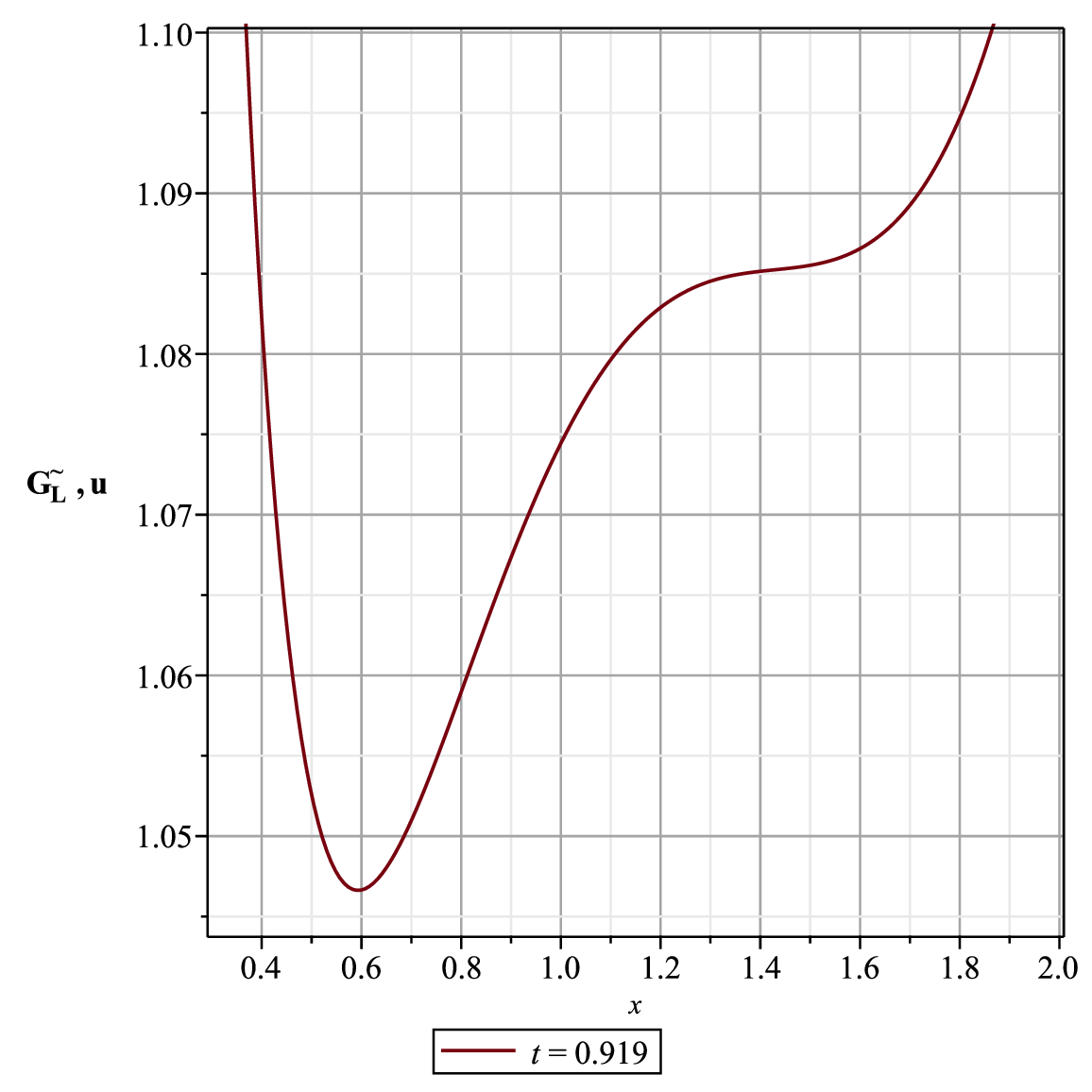}
 \label{11b}}
 \subfigure[]{
 \includegraphics[height=4.20cm,width=4cm]{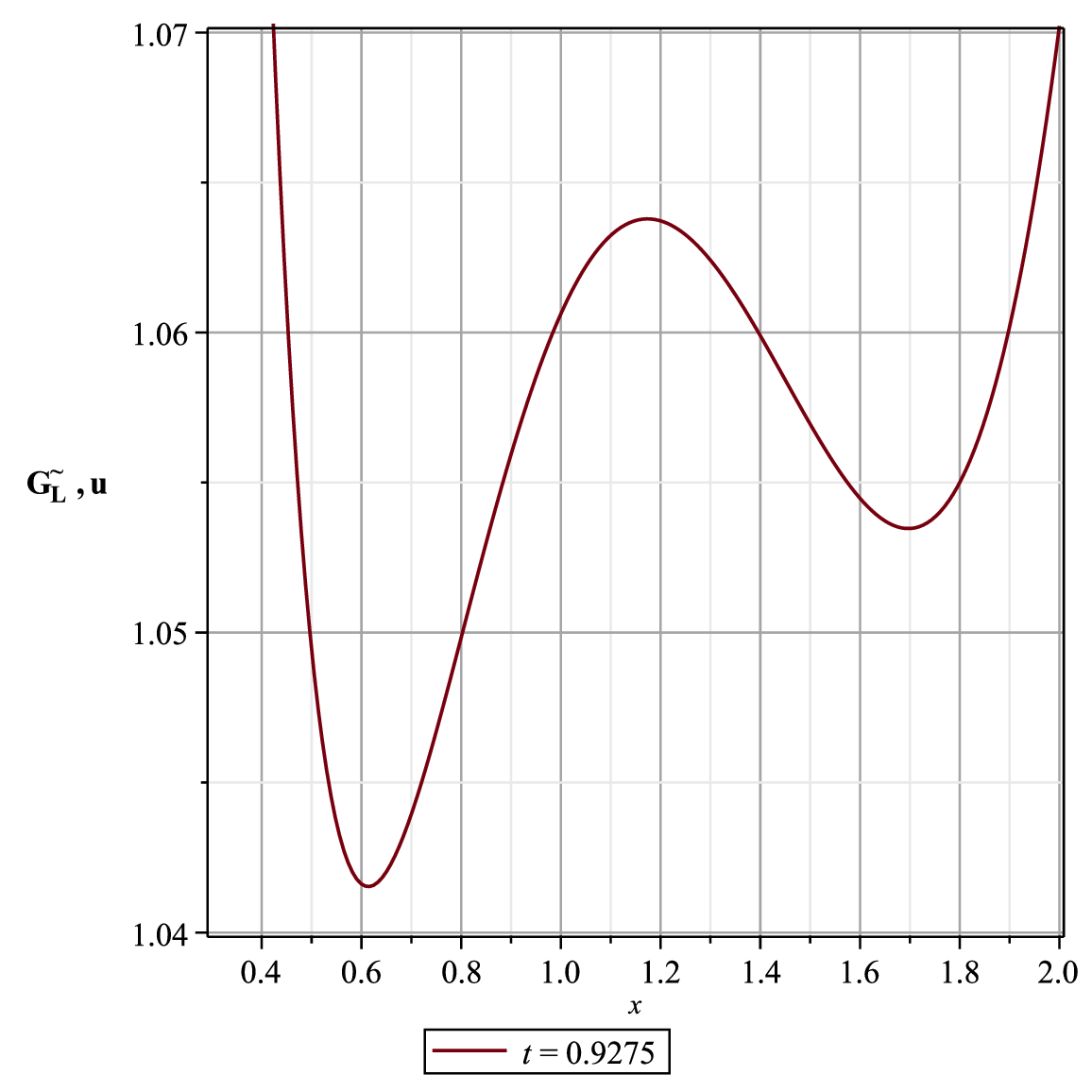}
 \label{11c}}
 \subfigure[]{
 \includegraphics[height=4.20cm,width=4cm]{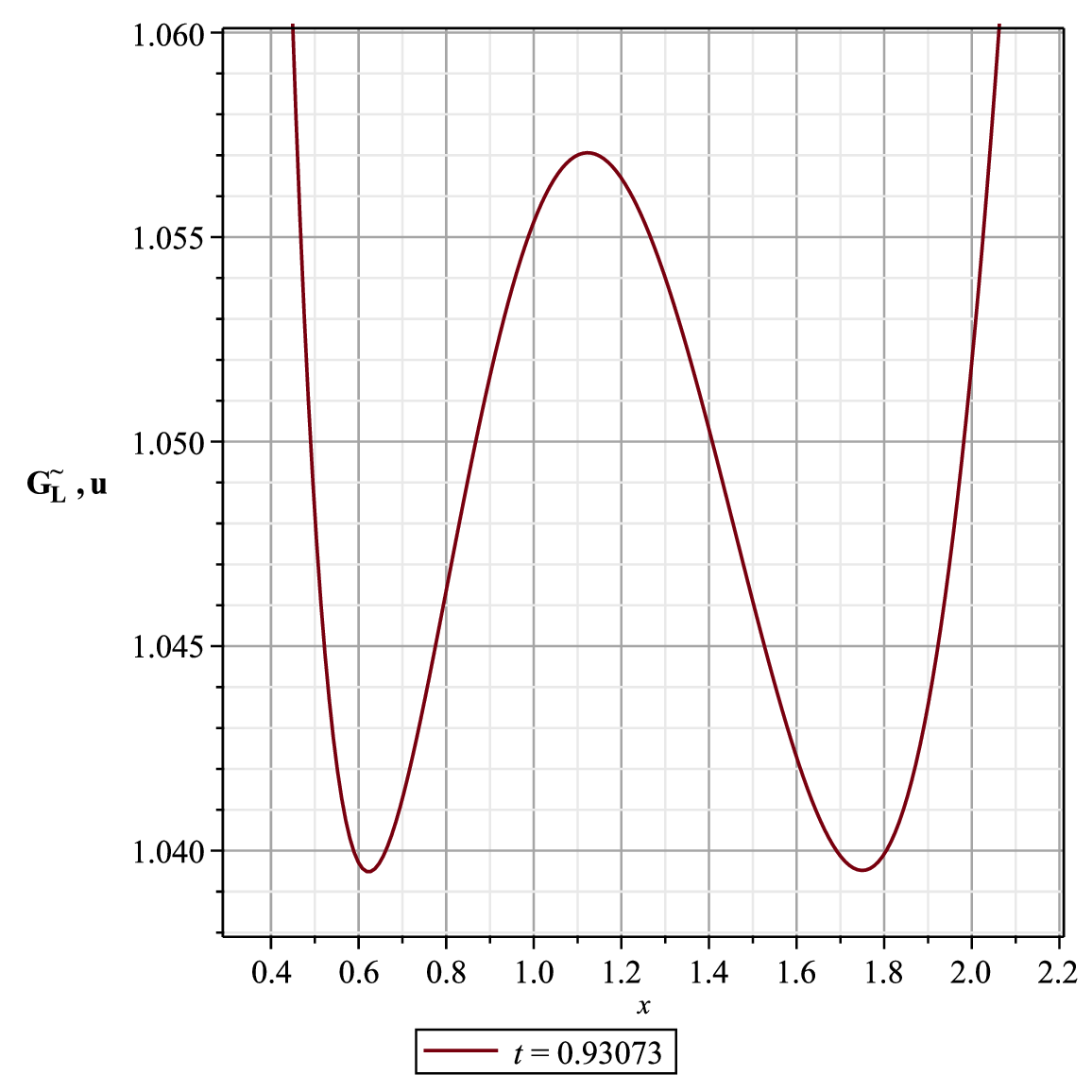}
 \label{11d}}
 \subfigure[]{
 \includegraphics[height=4.20cm,width=4cm]{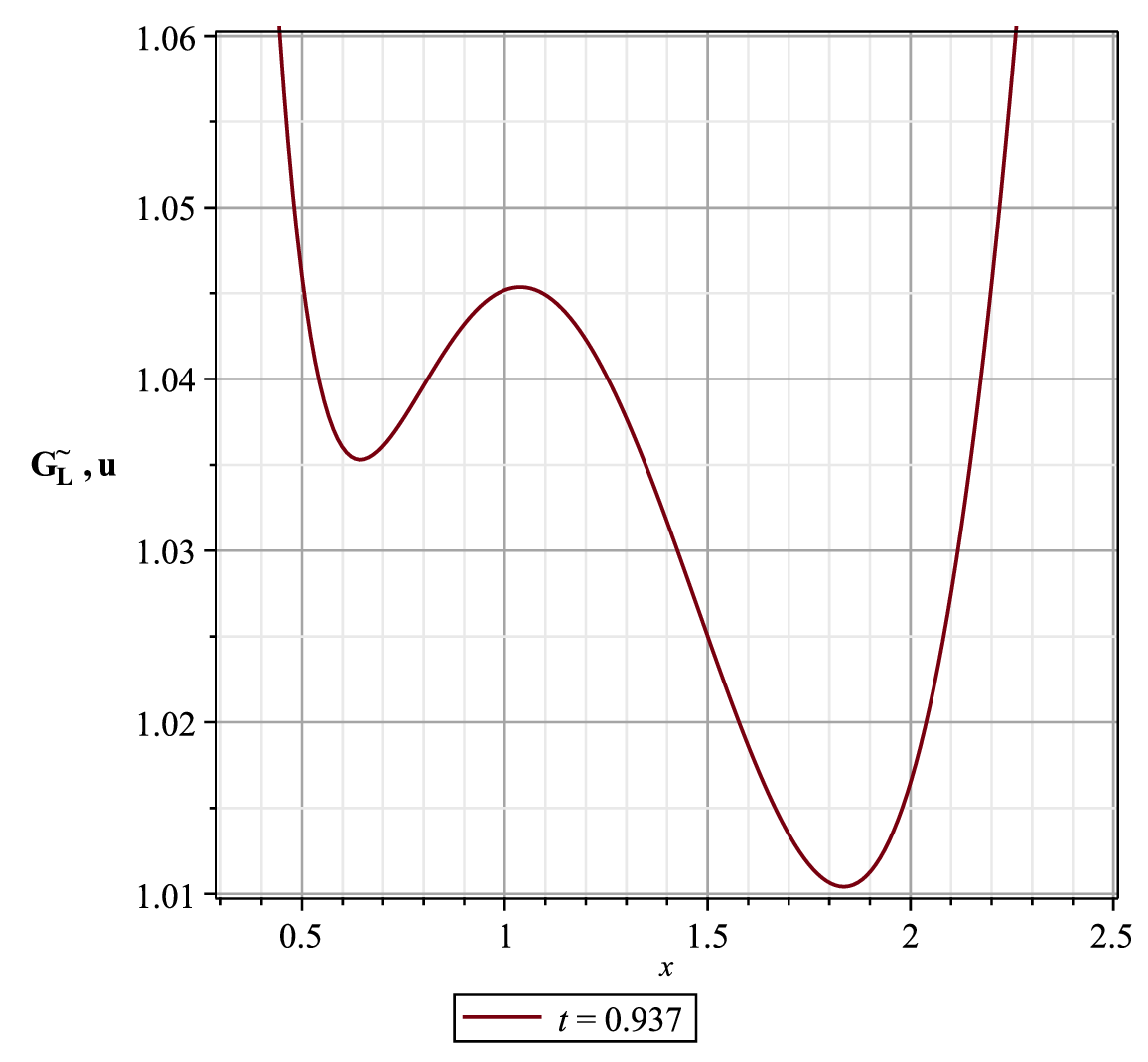}
 \label{11e}}
 \subfigure[]{
 \includegraphics[height=4.15cm,width=4cm]{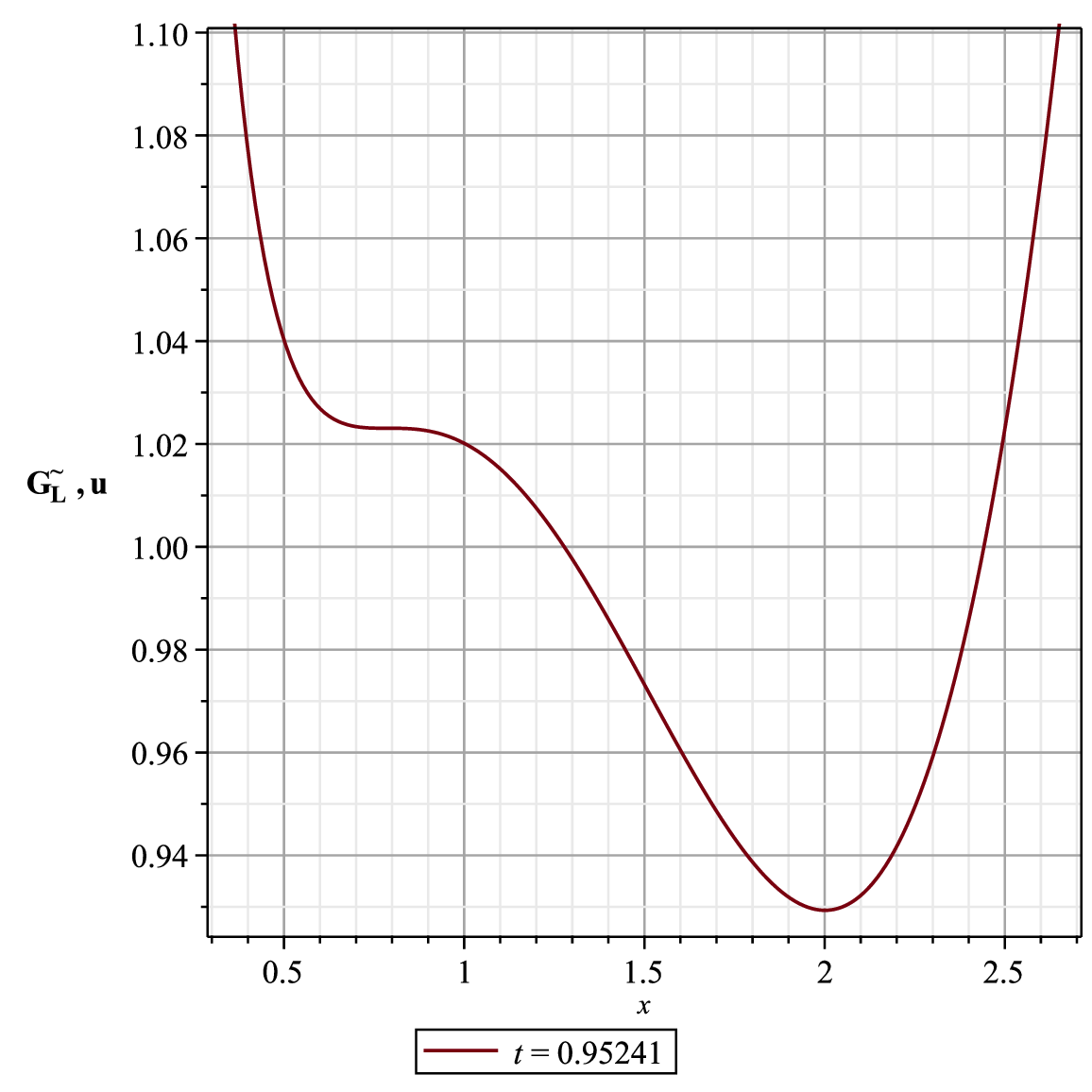}
 \label{11f}}
 \caption{\small{Behavioral sequence of energy in terms of x for different temperatures}}
 \label{m11}
\end{center}
 \end{figure}

The swallow tail representation of the Gibbs free energy landscape in terms of t and the behavioral sequence of the potential function in terms of distance and temperature for the (EGB-YM-CS) black hole is shown in Figs. (\ref{m10}), (\ref{m11}).

\end{document}